\documentclass[11pt,letter]{article}
\usepackage{amsfonts,latexsym,eucal,color,graphicx}
\usepackage{epsfig,amssymb}
\usepackage{graphicx}
\definecolor{red  }{rgb}{1,0,0}
\definecolor{blue }{rgb}{0,0,1}
\definecolor{green}{rgb}{0,1,0}
\begin{document}
% \draft

\begin{titlepage}

\begin{flushright}
{WU-AP/254/06, UTAP-567}
%%{hep-th/0609046}
\end{flushright}
\vspace{2.5cm}

%%%%%%%%%%%%%%%%%%%%%%%%%%%%%%%%%%%%%%%%%%%%%%%%%%%%%%%%%%%%%%%%%%%%
\begin{center}
	{\Large
		{\bf 
		New	stable phase of non-uniform charged black strings
		}
	}
\end{center}
\vspace{1cm}
%%%%%%%%%%%%%%%%%%%%%%%%%%%%%%%%%%%%%%%%%%%%%%%%%%%%%%%%%%%%%%%%%%%

\begin{center}
Umpei Miyamoto$^{a,1}$ and Hideaki Kudoh$^{b,2}$

%%--------   Address  ------------------
\vspace{1.0cm}
{\small {\textit{$^{a}$
Department of Physics,
 Waseda University, Okubo 3-4-1, Tokyo 169-8555, Japan, 
}}
}
\\
\vspace{5mm}
{\small \textit{$^{b}$
    Department of Physics, UCSB, Santa Barbara, CA 93106, USA
\\
    and
\\
Department of Physics, The University of Tokyo, Tokyo 113-0033, Japan
}}
\\
%%----------------------------------------
\vspace*{1.0cm}
%%----------- Email  ----------------------
{\small
{\tt{
$^{1}$ umpei at gravity.phys.waseda.ac.jp
\\
\hspace{0.7mm}
$^{2}$ kudoh at utap.phys.s.u-tokyo.ac.jp
}}
}
%%----------------------------------------
\end{center}

\vspace*{1.0cm}

%%%%%%%%%%%%%%%%%%%%%%%%%%%%%%%%%%%%%%%%%%%%%%%%%%%%%%%%%%%%%%%%%%%%%%%%
%%%%%%%%%%%%%%%%%%%%%%%%%%%%%%%%%%%%%%%%%%%%%%%%%%%%%%%%%%%%%%%%%%%%%%%%

%\date{\today}
%\preprint{2006-09-12, WU-AP/254/06, UTAP-***, hep-th/06*****}
%%%%%%%%%%%%%%%%%%%%%%%%%%%%%%%%%%%%%%%%%%%%%%%%%%%%%%%%%%%%%%%%%%%%%%%%

\begin{abstract}
Non-uniform black strings coupled to a gauge field are constructed by a perturbative method in a wide range of spacetime dimensions. 
At the linear order of perturbations, we see that the Gregory-Laflamme instability vanishes at the point where the background solution becomes thermodynamically stable.
The emergence/vanishing of the static mode resembles phase transitions, and in fact we find that its critical exponent is nearly 1/2, which means a second-order transition. 
By employing higher-order perturbations, the physical properties of the non-uniform black strings are investigated in detail. 
For fixed spacetime dimensions, we find the critical charges at which the stability of non-uniform states changes. 
For some range of charge, non-uniform black strings are entropically favored over uniform ones. 
The gauge charge works as a control parameter that controls not only the stability of uniform black strings but also the non-uniform states. 
In addition, we find that for a fixed background charge the uniform state is not necessarily the state carrying the largest tension. 
The phase diagram and a comparison with the critical dimension are also discussed.
\end{abstract}

\end{titlepage}

%%%%%%%%%%%%%%%%%%%%%%%%%%%%%%%%%%%%%%%%%%%%%%%%%%%%%%%%%%%%%%%%%%%%%%%%
\section{Introduction}
%%%%%%%%%%%%%%%%%%%%%%%%%%%%%%%%%%%%%%%%%%%%%%%%%%%%%%%%%%%%%%%%%%%%%%%

It is known that black objects with translational symmetries such as black branes and black strings suffer from the Gregory-Laflamme (GL) instability, breaking the translational symmetries~\cite{Gregory:1993vy,Gregory:1994bj}.
An expectation is that they decay into individual black holes, having larger entropy.
This naive expectation, however, was called into question because of the result by Horowitz and Maeda~\cite{Horowitz:2001cz}, which demonstrated that a black-string horizon cannot pinch off in finite affine time on the horizon. Although a dynamical simulation was reported~\cite{Choptuik:2003qd}, the endpoint of the collapse remains an outstanding open problem, and several possibilities have been pointed out~\cite{Marolf:2005vn,Garfinkle:2004em}.

As a solid approach, there are extensive studies of static black objects in compactified spacetimes, such as Kaluza-Klein (KK) spacetimes.
First, a non-uniform black string (NUBS) is constructed perturbatively in $ D=5 $ (five-dimensional) vacuum spacetime~\cite{Gubser:2001ac}, which was generalized to arbitrary dimensions and the critical dimension was discovered~\cite{Sorkin:2004qq,Kol:2004pn,Kudoh:2005hf}: the order of phase transition from uniform to non-uniform black strings is second order for $D \geq 14$ ($ D\geq 13$) in a microcanonical (canonical) ensemble, while it is first order for $D \leq 13$ ($D \leq 12$).
Recently, such critical dimensions were reproduced by an approach based on the Landau theory of phase transition~\cite{Kol:2006vu,Kol:2002xz}. Regarding the localized black holes in KK spacetimes, a perturbation method was developed, and small localized black holes were constructed~\cite{Harmark:2003yz,Gorbonos:2004uc,Gorbonos:2005px,Chu:2006ce,Karasik:2004ds}. 
After these perturbative analyses, the phase diagram of a black-hole and black-string system in fully non-linear regime was clarified in a series of papers~\cite{Wiseman:2002zc,Kudoh:2003ki,Kudoh:2004hs} (see also \cite{Kleihaus:2006ee} for improved numerical analysis for black string branch). The result suggests that the branches of static black hole and black string merge at a topologically changing solution. (See recent numerical work \cite{Kleihaus:2006ee, Sorkin:2006wp}.)

Although the GL instability is inevitable for neutral black strings/branes, charges can prevent them from suffering from the instability.
On this point, there is an interesting conjecture called the Correlated Stability Conjecture (CSC) (or called Gubser-Mitra conjecture~\cite{Gubser:2000ec,Gubser:2000mm}), which asserts that the GL instability of black objects with a non-compact translational invariance occurs iff they are (locally) thermodynamically unstable. This conjecture has been supported by several examples~\cite{Hirayama,Kang:2004hm,Gubser:2004dr}.
Although we speculate that the charges (e.g., a charge associated with a gauge field, KK momentum, angular momentum, and so on) play crucial roles in perturbative and non-linear regimes, our understanding of the stability and phase structure of charged black strings/branes is restricted to very special cases~\cite{Harmark:2004ws,Kudoh:2005hf,Hovdebo:2006jy}, in which the systems can be translated to a vacuum system.
The aim of this paper is to see the non-linear effects of a charge which stabilizes the uniform states at the linear order, as the CSC asserts.
For this aim, we perform the higher-order static perturbations of magnetic black strings.
And then, the physical properties of constructed solutions are investigated.
The main result is that for fixed spacetime dimensions there exist critical charges at which the order of phase transition changes in the sense that charged NUBSs become stable, compared with the uniform ones.
In other words, it means that the stable inhomogeneous black strings are allowed to exist in each dimension.
This is in contrast to the vacuum case, in which such stable states emerge only in higher dimensions ($D\ge 14$).

The organization of this paper is as follows.
In Sec.~\ref{sec:bg} we set up the problem by introducing an action integral, background solutions, and so on.
In Sec.~\ref{sec:linear} the linear perturbation is analyzed to some extent.
There, an interesting application of Landau theory of phase transition to charged system is discussed. In Sec.~\ref{sec:perturbation} we describe the higher-order perturbations in detail. 
With the numerical results of Secs.~\ref{sec:linear} and \ref{sec:perturbation}, we discuss the physical properties of charged NUBSs in Sec.~\ref{sec:physical properties}. 
Section~\ref{sec:conclusion} is devoted to summary and discussion.
Appendix~\ref{appendix:Ricci} includes the general form of explicit Einstein equations. 
We only consider magnetically charged black strings in this paper.
The possible other components, corresponding to an electric field, are discussed in Appendix~\ref{appendix:electric}. 
The calculation of some thermodynamic quantities are described in Appendices~\ref{appendix:capacity} and~\ref{appendix:entropy-diff}.

%%%%%%%%%%%%%%%%%%%%%%%%%%%%%%%%%%%%%%%%%%%%%%%%%%%%%%%%%%%%%%%%%%%%%%%%%%%%%%
\section{Setup}
\label{sec:bg}
%%%%%%%%%%%%%%%%%%%%%%%%%%%%%%%%%%%%%%%%%%%%%%%%%%%%%%%%%%%%%%%%%%%%%%%%%%%%%%
\subsection{Action and background solution}
%%%%%%%%%%%%%%%%%%%%%%%%%%%%%%%%%%%%%%%%%%%%%%%%%%%%%%%%%%%%%%%%%%%%%%%%%%%%%
We consider the following $(d+1)$-dimensional action ($d\geq 4$):
\begin{eqnarray}
I_{d+1}
	=
	\frac{1}{16\pi G_{d+1}} \int d^{d+1}x \sqrt{-g}
		\left[
			R-\frac{1}{2(d-2)!} \mathcal{F}_{d-2}^{2}
		\right],
\label{eq:d+1-action}
\end{eqnarray}
where $G_{d+1}$ is a ($d+1$)-dimensional gravitational constant and $\mathcal{F}_{d-2}$ is a ($d-2$)-form field.
We denote the total spacetime dimensions by $D \equiv d+1$.
The equations of motion (EOMs), obtained by varying the action, are
\begin{eqnarray}
&&
	R_{\mu\nu} 
	 =
	\frac{1}{2(d-3)!} {\mathcal{F}}_{\mu}^{ \;\;\mu_2 \ldots \mu_{d-2} }
	{\mathcal{F}}_{ \nu \mu_2 \ldots \mu_{d-2} } - \frac{ d-3 }{ 2(d-1)! }
	g_{ \mu\nu } {\mathcal{F}}^{2},
\cr
&&
	\partial_{\mu}
	\left(
		\sqrt{-g}\;{\mathcal{F}}^{ \mu\mu_{2} \ldots \mu_{d-2} }
	\right)
	 = 0.
\label{eq:form-EOM}
\end{eqnarray}
In addition, the form field satisfies the Bianchi identity, $\mathrm{d}{\mathcal{F}_{d-2}}=0$.

In this paper, we find non-uniform black string solutions emerging from the GL critical point.
Given the fact that the GL critical mode is a $s$-wave one~\cite{Kudoh:2006bp}, it will be convenient to begin with the following static axisymmetric metric ansatz:  
\begin{eqnarray}
	ds^2_{d+1} 
	= 
	- e^{2A(r,z)} dt^2 + e^{2B(r,z)} 
	\left[ e^{2H(r)}dr^2 + dz^2\right]
	+ r^2 e^{2C(r,z)} d\Omega^2_{d-2}.
\label{eq:conformal}
\end{eqnarray}
Here, we assume that the horizon of black string extends along the $z$-direction.
For uniform black strings, the functions $A, B, C$ depend only on the radial coordinate $r$. 
Although the function $H(r)$ can be eliminated by a radial gauge transformation, we keep $H(r)$ for later convenience and generality.
The Ricci tensor and explicit EOMs for this metric ansatz are included in Appendix \ref{appendix:Ricci}.
With this metric ansatz, we can find a general solution of magnetic field by solving Bianchi identities:
\begin{eqnarray}
    {\mathcal{F}} = Q_m ~ \varepsilon_{d-2}, 
\label{eq:general-F}
\end{eqnarray}
where $Q_m$ is a constant and $\varepsilon_{d-2} $ is the volume element of a unit $(d-2)$-sphere.
In the rest of this paper, we only consider magnetic solutions.
Thus, the general solution~(\ref{eq:general-F}) of the form fields makes our analysis very simple; 
we do not need to consider the perturbations of the form field independently.
For $d=4$ and $d=5$, the gauge field can carry an electric charge.
In particular, the $3$-form field becomes self-dual in $d=5$ and the electric field gives the same results as those for the magnetic case (Appendix~\ref{appendix:electric}).

By the dimensional reduction method, the uniform black string and black brane solutions of the action~(\ref{eq:d+1-action}) can be constructed from the black hole solutions in a $d$-dimensional dilatonic system~\cite{Horowitz:1991cd}.
The explicit expression of a magnetically charged black string is given by
\begin{eqnarray}
&&
ds_{d+1}^2=
	-f_{+}dt^2
	+\frac{1}{f_{+}f_{-}}dr^{2}
	+f_{-}dz^2
	+r^2  d\Omega_{d-2}^{2},
\nonumber
\\
&&
f_{\pm}(r) = 1 -\left(\frac{r_{\pm}}{r}\right)^{d-3},
\nonumber
\\ 
&& 
{\mathcal{F}} = 
	\sqrt{(d-1)(d-3)}\, (r_{+}r_{-})^{(d-3)/2} \, \varepsilon_{d-2} . 
\label{eq:UBSmetric}
\end{eqnarray}
For $0 < $ $r_{-} < r_{+}$, this solution has an event horizon at $r=r_{+}$ and an inner horizon at $r=r_{-}$ where a curvature singularity exists. 
Introducing new variables by
\begin{eqnarray}
&&
	A = a + \ln \sqrt{f_+}, \quad
	B = b + \ln \sqrt{f_{-}}, \quad
	C = c, \quad
	H = -\frac{1}{2} \ln \left[ f_{+} f_{-}^{2} \right] ,
\label{eq:conformal-to-string}
\end{eqnarray}
a deformed black string can be described by
\begin{eqnarray}
&&
ds_{d+1}^2=
	- e^{2a(r,z)}f_{+}dt^2
	+ e^{2b(r,z)}f_{-}
		 \left[
			\frac{ dr^{2} }{f_{+}f_{-}^2}
			+ dz^2
		 \right]
	+ e^{2c(r,z)}r^2  d\Omega_{d-2}^{2}.
\label{eq:string}
\end{eqnarray}
Substituting the transformation (\ref{eq:conformal-to-string}) into the general form of Ricci tensor (\ref{eq:ricci}) and source term (\ref{eq:Einstein-source}), we obtain EOMs for $a$, $b$ and $c$.

%%%%%%%%%%%%%%%%%%%%%%%%%%%%%%%%%%%%%%%%%%%%%%%%%%%%%%
\subsection{Perturbation scheme}
\label{sec:pert-scheme}
%%%%%%%%%%%%%%%%%%%%%%%%%%%%%%%%%%%%%%%%%%%%%%%%%%%%%%

Following \cite{Gubser:2001ac}, we can construct non-uniform black strings/branes by static perturbations, as well as to specify the GL critical point quite easily.
Before the detailed discussion, which will be given in Secs.~\ref{sec:linear} and \ref{sec:perturbation}, let us introduce the general perturbation scheme here.

First, we rescale coordinates $r$, $z$, and ``charge" parameter $r_{-}$ by the horizon radius:
\begin{eqnarray}
	y \equiv \frac{r}{r_+},
		\;\;
	x \equiv \frac{z}{r_+},
		\;\;
	q \equiv \frac{r_-}{r_+}.
\end{eqnarray}
Then, we expand the metric function $X(x, y)$ ($X=a, b, c$) around the uniform solution,
\begin{eqnarray}
 &&
	X (x,y) = 
		\sum_{n=0}^{\infty} \epsilon^n X_n(y) \cos (n K x),
\nonumber
 \\
 &&
	X_n(y) = \sum_{p=0}^{\infty} \epsilon^{2p} X_{n,p}(y), 
		\;\;\;
	K  = \sum_{q=0}^{\infty} \epsilon^{2q} k_{q},
\label{eq:Def Expansion}
\end{eqnarray}
where $X_{0, 0}(y)=0$ is imposed.
Here $K$ is the GL critical wavenumber, in other words, $L\equiv 2\pi/K$ gives the asymptotic length of the compactified space, and
$\epsilon$ is an expansion parameter.
Substituting these expansions into the Einstein equations, we obtain ordinary differential equations (ODEs), to be solved order-by-order (Table \ref{table:X_nm}).

%%%%%%%%%%%%%%%%%%% Table %%%%%%%%%%%%%%%%%%%%%%%%%%%%%%%%%%%%
\begin{table}[tb]
\begin{center}
\begin{tabular}{c|lll|cc}
\hline  \hline 
               & Zero     &~~& KK          & $K$       
\\  \hline 
$O(\epsilon)$  &          &&$X_{1,0}$      &  $k_0\sim O(1)$  
\\
 $O(\epsilon^2)$&$X_{0,1}$&&$X_{2,0}$      &      
\\
 $O(\epsilon^3)$&         &&$X_{3,0}$, ~ $X_{1,1}$& $k_1 \sim O(\epsilon^2) $
\\ \hline  \hline
\end{tabular}
\caption[short]{
Table of unknown functions up to the 3rd order.
A mode $X_{n,p}$ appears at $O(\epsilon^{n+2p})$. The zero modes $X_{0,p}$ appear at $O(\epsilon^{2p})$.
In the last column, the wavenumber $k_0$ of order $O(1)$ and its corrections are listed to show at which order they enter into the decoupled ODEs. 
}
\label{table:X_nm}
\end{center}
\end{table}
%%%%%%%%%%%%%%%%%%% Table %%%%%%%%%%%%%%%%%%%%%%%%%%%%%%%%%%%%%

To have an insight into the whole perturbation analysis, it would be nice to see the linear perturbations.
For simplicity, let us focus on $d=5$, and we always restrict ourselves to this number of dimensions when we write down perturbation equations hereafter. 
The structures of EOMs and boundary conditions are similar for the other dimensions (see Eqs. (\ref{eq:ricci}) and (\ref{eq:Einstein-source})).
At the linear order $O(\epsilon)$, we have the $X_{1,0}$ mode. 
By combining Einstein equations, we obtain coupled ODEs, which determines $a_{1,0}(y)$ and $c_{1,0}(y)$:
\begin{eqnarray}
 &&
	\widehat{L}_{k_0}^{[1]}a_{1,0}+\widehat{P}^{[1]}c_{1,0}=0,
 \nonumber
 \\
 &&
 	\widehat{L}_{k_0}^{[2]}c_{1,0}+\widehat{P}^{[2]}a_{1,0}=0,
\label{eq:EOM-X10}
\end{eqnarray}
where $\widehat{L}_{k_0}^{[i]}$ and $\widehat{P}^{[i]}$ ($i=1,2$) are linear operators given by
\begin{eqnarray}
 &&
 	\widehat{ L }_{k_0}^{[1]} =	\frac{d^2}{dy^2}
		+ \frac{6y^4-9y^6+q^2(-8+8y^2+3y^4)}{y(q^2-y^2)(2-5y^2+3y^4)}
		  \frac{d}{dy}
 		+
	\frac{4q^4-4q^2+k_0^2y^6(2-3y^2)}{(q^2-y^2)^2(2-5y^2+3y^4)},
 \nonumber
 \\
 &&
	\widehat{L}_{k_0}^{[2]}=\frac{d^2}{dy^2}
		+ \frac{2[-4y^2+6y^4-3y^6+q^2(8-13y^2+6y^4)]}{y(q^2-y^2)(2-5y^2+3y^4)}
		  \frac{d}{dy}
 \nonumber
 \\
 &&
 \hspace{7cm}
	+ \frac{12q^4+4y^4+2k_0^2y^6-3k_0^2y^8+4q^2(3-10y^2+3y^4)}{(q^2-y^2)^2(2-5y^2+3y^4)},
 \nonumber
 \\
 &&
 \widehat{P}^{[1]}
	=	\frac{3[2y^2-3y^4+q^2(-6+7y^2)]}{y(q^2-y^2)(2-5y^2+3y^4)}\frac{d}{dy}
	   - \frac{12q^2(1+q^2-2y^2)}{(q^2-y^2)^2(2-5y^2+3y^4)},
 \nonumber
 \\
 &&
 \widehat{P}^{[2]}
	=	\frac{(-1+y^2)[3q^2(-2+y^2)+y^2(2+y^2)]}{y(q^2-y^2)(2-5y^2+3y^4)}\frac{d}{dy}
		-\frac{4[q^4+y^4-q^2(1+y^4)]}{(q^2-y^2)^2(2-5y^2+3y^4)}.
\label{eq:operators}
\end{eqnarray}
The function $b_{1, 0}(y)$ is not independent but can be written in terms of $a_{1, 0}(y)$ and $c_{1, 0}(y)$,
\begin{eqnarray}
 b_{1, 0}(y) 
 = 
 \frac{ (-1+q^2) y^2 a_{1,0} + (-1+y^2) [ (6q^2-3y^2) c_{1,0} + y(q^2-y^2)(a_{1,0}+3c_{1,0}) ]}
      {(q^2-y^2)(-2+3y^2)}.
\label{eq:b10}
\end{eqnarray}
As we will discuss in Sec.~\ref{sec:linear}, Eq.~(\ref{eq:EOM-X10}) can be solved as a shooting problem.
The equations for higher-order perturbations can be reduced to similar systems of ODEs.

%%%%%%%%%%%%%%%%%%%%%%%%%%%%%%%%%%%%%%%%%%%%%%%%%%
\subsection{Physical quantities}
\label{sec:Physical quantities}
%%%%%%%%%%%%%%%%%%%%%%%%%%%%%%%%%%%%%%%%%%%%%%%%%%

Here, we prepare the expressions for physical quantities of the metric (\ref{eq:string}), with which we will later discuss the thermodynamics of non-uniform black strings.

To calculate asymptotic charges such as mass and tension, one has to know the asymptotic behavior of the metrics. 
As we will see in the following sections, homogeneous perturbations (zero modes) appear at the second order, and they decay according to some power law in the asymptotic region $r\gg r_{+}$. 
Suppressing all exponentially small corrections, which comes from Kaluza-Klein (KK) modes in the first- and second-order perturbations, we find that the leading asymptotic behaviors of the metric functions for $d \geq 5$ are~\footnote{For $d=4$, the asymptotic form of $c(r,z)$ is exceptional: $ c(r,z) \simeq C_{\infty} (r_{+}/r) \ln (r/r_+) $.}
\begin{eqnarray}
 && 
 	a(r,z) 
 	\simeq 
 	A_{\infty} \left( \frac{ r_+ }{ r } \right)^{ d-3 },
 \nonumber\\
 && b(r,z)
 	\simeq
 	B_{\infty} \left( \frac{ r_+ }{ r } \right)^{ d-3 },
 \label{eq:asymptotics}
 \nonumber
 \\
 && 
	c(r,z)
 	\simeq 
	C_1 \frac{r_+}{r} + C_{\infty}\left(\frac{r_{+}}{r}\right)^{d-3},
 \label{eq:asymptotics}
\end{eqnarray} 
where $A_{\infty}$, $B_{\infty}$, $C_{\infty}$ and $C_1$ are constants.
From the Einstein equations in the asymptotic region, these constants are related as
\begin{eqnarray}
	A_{\infty} + 2 B_{\infty} + (d-4) C_{\infty} = 0.
\end{eqnarray}
With these asymptotics, the mass per unit length $M/L$ and relative tension $n$ (and tension $\mathbb{T}$) in the $z$-direction ($z\in[0,L]$) for $d \geq 5$ are computed~\cite{Hawking:1995fd,Harmark:2003dg} as
\begin{eqnarray}
 &&
	\frac{M}{L} 
 		= 
		\frac{ \Omega_{d-2} r_+^{d-3} }{ 16 \pi G_{d+1} } 
  			\left[
 				2(2d-5) B_{\infty} + 2(d-2)(d-4)C_{\infty} + d-2 + q^{d-3}
  			\right],
  	\label{eq:mass}
 \\
 &&
	 n
 		 \equiv
		\frac{L \mathbb{T}}{M}
 		 =
 		\frac{  1 -2 (d-4) B_\infty + 2(d-4) C_\infty + (d-2) q^{d-3}}
              {
                 (d-2)[ 1 + 2 (d-4) C_\infty ] + 2(2d-5) B_\infty + q^{d-3}
               },
 	\label{eq:tension}
\end{eqnarray}
where $\Omega_{d-2} = 2 \pi^{ (d-1)/2 } / \Gamma[ (d-1)/2 ]$ is the surface area of a unit ($d-2$)-sphere.

Entropy $S$, temperature $T$, magnetic charge $Q$, and chemical potential $\Phi_{H}$ are given by
\begin{eqnarray}
 &&
	T
	=
	\frac{ d-3 }{ 4\pi r_+ } \sqrt{ f_- } ~ e^{a - b} \Big|_{ r=r_+ },
 \label{eq:temperature}
 \\
 &&
 	\frac{S}{L}
 	=
 	\frac{ \Omega_{d-2} }{ 4 G_{d+1} } r_+^{d-2} \sqrt{f_-} 
   		\langle e^{b+(d-2)c}  \rangle \Big|_{r=r_{+}},
 \label{eq:entropy}
 \\
 &&
	\frac{Q}{L}
	=
	\frac{ (d-1) \Omega_{d-2} r_+^{d-3} }{ 16 \pi G_{d+1} } q ^{ (d-3)/2 },
 \label{eq:charge}
 \\
 &&
	\Phi_{H}
	=
	(d-3) r_{+}^{d-3} q^{(d-3)/2}
		\int_{r_+}^{\infty}
		\frac{ dr }{ r^{d-2} }
		\langle
			e^{a+2b-(d-2)c}
		\rangle,
 \label{eq:potential}
\end{eqnarray}
where $\langle \;\cdot\; \rangle$ denotes an averaging in the $z$-direction.
Here, some remarks on these quantities may be needed.
Although the $z$-independence of the temperature (corresponding to the zeroth law) is not obvious from the expression~(\ref{eq:temperature}), it will be verified from the boundary conditions which state that $a-b$ for all modes vanishes on the horizon except for the homogeneous mode. The magnetic charge~(\ref{eq:charge}) is normalized so that  we have $Q/M \to 1$ in the extremal limit $q\to1$. 
The chemical potential~(\ref{eq:potential}) can be calculated in two ways. One is to use the thermodynamical relation $ \Phi_H = (\partial M/\partial Q)_{S} $. For uniform black strings, this method is demonstrated in Appendix~\ref{appendix:capacity}.
The second way is based on the electric potential, by taking the dual field $\ast {\mathcal{F}}$ of the field ${\mathcal{F}}$. 
We integrate out the dual field along the $z$-direction, by the KK reduction, and introduce the electric potential on the horizon by  $\Phi_H = \mathcal{A} |_{r=r_+}$, where $\ast {\mathcal{F}} = \mathrm{d} \mathcal{A}$.

According to the correlated stability argument~\cite{Gubser:2000ec,Gubser:2000mm}, the dynamical stability of the background spacetime is closely related to its thermodynamical stability. 
In our case, the thermodynamical stability is determined by the sign of a specific heat, given by (Appendix \ref{appendix:capacity})
\begin{eqnarray}
 	C_{Q} 
 	\equiv
 	\left(
 		\frac{ \partial M }{ \partial T }
 	\right)_{Q}
 	=
 	\frac{ L \Omega_{d-2} r_+^{d-2} }{ 4 G_{d+1} }
 	\frac{ \sqrt{ 1-q^{d-3} } [(d-2) - q^{d-3}] }{ -1 + (d-2)q^{d-3}}.
 \label{eq:def capacity}
\end{eqnarray}
The specific heat Eq.~(\ref{eq:def capacity}) is negative for a small charge, but it becomes positive at a critical charge $Q_{\mathrm{GM}}$, which is given parametrically  by $q=q_{\mathrm{GM}}$,
\begin{eqnarray}
    q_{\mathrm{GM}} = \frac{1}{(d-2)^{1/(d-3)}}.
 \label{eq:qc}
\end{eqnarray}
Thus the system will be stable for $ Q > Q_{\mathrm{GM}} $.
As we will see below, the GL critical mode indeed disappears at this point, and the correlated stability is realized.

Note that the choice of thermodynamic ensemble depends on crucially on whether we treat the charge as a parameter that has been fixed or as a parameter that can vary freely~\cite{Hirayama:2002hn,Ross:2005vh,Kudoh:2005hf,Harmark:2005jk}.
The above condition on the thermodynamic stability, $C_Q>0$, is for a canonical ensemble, in which the temperature and charge are kept fixed.
If the solution is \textit{smeared}, in a magnetic case this means the string is charged along a transverse direction, the charge itself can redistribute freely in the smeared direction, and we have to consider the stability in a grandcanonical ensemble in which the stability condition is specified by the specific heat and isothermal permittivity.

\section{Static linear perturbation : $X_{1,0}$}
\label{sec:linear}
%%%%%%%%%%%%%%%%%%%%%%%%%%%%%%%%%%%%%%%%%%%%%%%%%%%%%%

%%%%%%%%%%%%%%%%%%%%%%%%%%%%%%%%%%%%%%%%%%%%%%%%%%%%%%
\subsection{Search for static mode}
\label{sec:static-mode}
%%%%%%%%%%%%%%%%%%%%%%%%%%%%%%%%%%%%%%%%%%%%%%%%%%%%%%

We can specify the critical wavenumber $k_0$ for a given value of $q$ by solving the coupled ODEs (\ref{eq:EOM-X10}) with suitable boundary conditions.
Solving the EOMs in the asymptotic region ($y\gg 1$), we see that the perturbations resulting in regular solutions must have the following asymptotic behaviors:
\begin{eqnarray}
	a_{1,0} ,~ c_{1,0} \sim  e^{-k_0 y}.
\label{eq:asymptotics-X10}
\end{eqnarray}
From these asymptotics, it is understood that $a_{1,0}(y)$ and $c_{1,0}(y)$ are the KK modes, which are localized near the horizon.

The event horizon is a regular singular point of the differential equations.
Demanding the regularity of the perturbations on the horizon $(y=1)$, the following boundary conditions are required (for $d=5$):
\begin{eqnarray}
 &&
	a_{1, 0}^{\prime}|_{y=1}
	=
	\frac{
	 	 2(k_0^2 -6 + 16q^2 - 18q^4 ) a_{1,0} 
	 	-3(k_0^2 -4  + 24q^2 - 20q^4 ) c_{1,0} 
	     }
	{ 6(1-q^2)^2},
 \nonumber
 \\
 &&
 	c_{1, 0}^{\prime}|_{y=1}
 	=
 	2a_{1,0} 
 	+ 
 	\frac{ -4 + k_0^2 + 16q^2 - 12q^4 }{ 2( 1-q^2 )^2 } c_{1,0} .
 \label{eq:regularity-X10}
\end{eqnarray}
Thus, these boundary conditions constrain the values of perturbations and their first derivatives.
Since we are working on linear perturbations and the amplitude of perturbations is specified by $\epsilon$, we are free to fix the amplitude of a function at any point.
Therefore, we set $c_{1, 0}(1)=1$.
The regularity at the horizon requires also $ b_{1,0}(1) = a_{1,0}(1) $, corresponding to the zeroth law of thermodynamics, as mentioned in Sec.~\ref{sec:Physical quantities}.

\begin{figure}[t]
	\begin{center}
	\includegraphics[width=8cm]{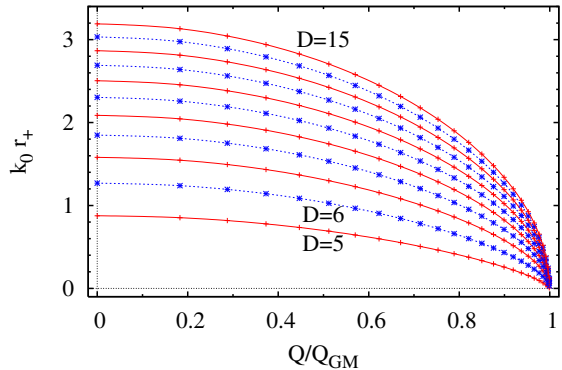}
		\caption{	
Charge dependence of GL critical wavenumber $k_{0}$, normalized by the inverse of the horizon radius, for $D=5\sim 15$. 
We see that the critical wavenumbers decrease monotonically as the background charge increases.
As expected from the correlated stability conjecture, these decreasing wavenumbers vanish at the critical charge $Q_{ \mathrm{GM} }$, above which the background uniform black strings are thermodynamically stable. 
\label{fg:k0-Q} }
	\end{center}
\end{figure}

Now, we are ready to solve (\ref{eq:EOM-X10}) for each $q$ with the boundary conditions (\ref{eq:asymptotics-X10}) and (\ref{eq:regularity-X10}). 
For the numerical method, we integrate (\ref{eq:EOM-X10}) from the horizon by taking $a_{1,0}(1)$ and $k_{0}$ as shooting parameters.
The shooting parameters resulting in regular solutions for the neutral case ($q=0$) are given in Tables~\ref{table:boundary-values} for several dimensions ($D=6, 10$ and $14$).
The charge dependence of the critical wavenumber is depicted in Fig.~\ref{fg:k0-Q}.
We see that $k_0$ decreases monotonically as the background charge increases, and then vanishes at the critical charge (\ref{eq:qc}), above which the specific heat of the background solution is positive. 
This is a clear realization of the correlated stability.

%%%%%%%%%%%%%%%%%%%%%%%%%%%%%%%%%%%%%%%%%%%%%%%%%%%%%%%%%%%%%%%%%%%%%%%%%%%%%%%
%%%%%%%%%%%%%%%%%%%%%%%%%%%%%%%%%%%%%%%%%%%%%%%%%%%%%%%%%%%%%%%%%%%%%%%%%%%%%%%

%%%%%%%%%%%%%%%%%%%%%%%%%%%%%%%%%%%%%%%%%%%%%%%%%%%%%%
\subsection{Transition}
\label{sec:phase-trans}
%%%%%%%%%%%%%%%%%%%%%%%%%%%%%%%%%%%%%%%%%%%%%%%%%%%%%%

The emergence of the static mode and its associated non-uniform states at $Q<Q_{\mathrm{GM}}$ looks like a phase transition in condensed matter physics.
\footnote{
We would like to thank Barak Kol, who pointed out this to us. 
} 
A simple example of phase transition is as follows.
At high temperatures, there is no order, and the ``order" parameter $\left\langle \phi \right\rangle$ is zero. At a critical temperature, $T_c$, order sets in so that, for temperatures below $T_c$, $\left\langle \phi \right\rangle$ is nonzero. If $\left\langle \phi \right\rangle$ rises continuously from zero, the transition is second order and the specific heat has a jump discontinuity at the transition.
\cite{Chaikin:1995book}. 
The same thing holds for the present case, by just replacing $\left\langle \phi \right \rangle$ and $T$ by $k_0$ and $Q$, respectively, although the ``order" (translational symmetry) sets in at $Q>Q_{\mathrm{GM}}$. The specific heat (\ref{eq:def capacity}) also has 
a discontinuity at $Q=Q_{\mathrm{GM}}$.
%%\footnote{ 
%%In consensed matter physics, the specific heat is positive at the ordered phase at $T<T_c$. 
%%But in the present case it is negative. 
%%In addition to the notion of "order", 
%%meaning of many idea must be "invesed". 
%%}
%%%
This kind of phase transition can be well described by the Landau phenomenological mean-field theory. 
The mean-field theory, in which the free energy takes the form of $f= r\left\langle \phi\right\rangle^2 +  u  \left\langle \phi\right\rangle^4$ with $u>0$, predicts a second-order phase transition with
\begin{eqnarray}
 \left\langle \phi \right\rangle \sim (T_c -T)^\beta, 
 \label{eq:critical exponent}
\end{eqnarray}
where $\beta=1/2$ is called the critical exponent. 
It is interesting to see if the emergence of the static mode follows the line of this argument. The critical exponents extracted from the results presented in Fig.~\ref{fg:k0-Q} are
\begin{eqnarray}
\beta =0.55 ~~ ({D=6}),
\qquad 
\beta =0.51~~({D=7}),
\qquad
\beta = 0.50 ~~({D \ge 8}),
\end{eqnarray}
These values are fairly close to the prediction of the Landau theory.
(Note that we can also analyze the critical exponent for the mass per unit length, i.e. $k_0 M/2\pi$, as an another physically motivated quantity. Such analysis gives the same result as above because the contribution of $q$ in $M$ is negligibly small.)
$\beta$ in $D\le 7$ deviates slightly from the prediction. 
The determination of $k_0$ near the GM point requires high numerical accuracy (in particular for smaller dimensions), and hence the derived critical exponent will contain numerical errors. Taking this into account, we are well justified in expecting that the critical exponent for $k_0 r_+$ (or $k_0 M/2\pi$) is exactly $\beta=1/2$ and that it will not depend on the dimensions. 
Moreover, it is an interesting issue to study the universality of this phenomenon.
(See \cite{Gubser:2002yi} for an earlier study of this kind of observation in other systems.)
However, it is beyond the scope of this paper, and we leave it for future study. 
Instead, we extend the perturbative analysis to higher orders and identify the physical properties of charged non-uniform states.

%%%%%%%%%%%%%%%%%%%%%%%%%%%%%%%%%%%%%%%%%%%%%%%%%%%%%%
\section{Static perturbations at higher orders}
\label{sec:perturbation}
%%%%%%%%%%%%%%%%%%%%%%%%%%%%%%%%%%%%%%%%%%%%%%%%%%%%%%

In this section, we discuss the structure of the EOMs, boundary conditions and asymptotics for higher-order perturbations.
For simplicity, we set the dimension to $d=5$ in the rest of this section, although the analyses are done for several spacetime dimensions. 
The dimensional dependence appears mainly in the asymptotic behavior for the zero modes at the second order. 
When we solve the ODEs at each order, a non-trivial task is to learn how to fix the integration constants. We will also discuss this point to some extent, 
in addition to other technical aspects.
Readers who are interested only in the physical results can jump to Sec.~\ref{sec:physical properties}\footnote{A Mathematica notebook including numerics in this paper will be available at the author's website, 
$\mathrm{http://www.gravity.phys.waseda.ac.jp/}\widetilde{~~}\mathrm{umpei}$.
}.

%%%%%%%%%%%%%%%%%%%%%%%%%%%%%%%%%%%%%%%%%%%%%%%%%%%%%%%%%%%%%%%%%%%%%%%%%%%%%%%
\subsection{Second-order perturbation}
%%%%%%%%%%%%%%%%%%%%%%%%%%%%%%%%%%%%%%%%%%%%%%%%%%%%%%%%%%%%%%%%%%%%%%%%%%%%%%%

At the second order $O(\epsilon^2)$, we have two independent modes, $X_{2,0}$ and $X_{0,1}$. 
The former is a second-order counterpart of $X_{1,0}$, which is straightforward to integrate. 
On the other hand, the latter is the lowest-order zero mode.
This mode has a different structure of EOMs and thereby the asymptotic behaviors are different from those for KK modes.

%%%%%%%%%%%%%%%%%%%%%%%%%%%%%%%%%%%%%%%%%%%%%%%%%%%%%%%%%%%%%%%%%%%%%%%%%%%%%%%
\subsubsection{KK mode: $X_{2,0}$}
%%%%%%%%%%%%%%%%%%%%%%%%%%%%%%%%%%%%%%%%%%%%%%%%%%%%%%%%%%%%%%%%%%%%%%%%%%%%%%%
The EOMs for the $X_{2,0}$ mode take the form of
\begin{eqnarray}
 &&
 	\hat{L}_{2k_0}^{[1]} a_{2,0}
 	 +
 	\hat{P}^{[1]}c_{2,0}
 	=
 	S_{2,0}^{[1]}( X_{1,0}; k_{0} ),
 \nonumber
 \\
 &&
 	\hat{L}_{2k_0}^{[2]} c_{2,0}
  	 +
  	\hat{P}^{[2]} a_{2,0} 
  	=
  	S_{2,0}^{[2]} ( X_{1,0}; k_{0} ),
 \label{eq:EOM-X20}
\end{eqnarray}
similar to $X_{1,0}$. 
Here, $S_{2,0}^{[i]}$ represents a source term, which is quadratic in $X_{1,0}(y)$ and its first derivative.
The function $b_{2,0}(y)$ is written in terms of $a_{2,0}(y)$, $c_{2,0}(y)$ and their first derivatives.
The asymptotic behaviors in the far region are
\begin{eqnarray}
	a_{2,0} 
	,~
	c_{2,0} \sim e^{ -2 k_0 y}.
 \label{eq:X20-asymptotics}
\end{eqnarray}
The horizon boundary conditions analogous to Eq.~(\ref{eq:regularity-X10}) can be immediately obtained from (\ref{eq:EOM-X20}): 
\begin{eqnarray}
&&	a_{2,0}^{\prime}(1) 
    =  \frac{1}{ 12(1-q^2)^2 } 		\Big[ 8( -3 + 2k_0^2 + 8q^2 - 5q^4 ) a_{2,0} 			-24( -1 + k_0^2 + 6q^2 - 5q^4) c_{2,0}
\nonumber
\\ &&
				\hspace{0.1cm}
			+2( -6 + k_0^2 + 16q^2 - 10q^4 ) a_{1,0}^2
			+3( 8 + k_0^2 - 48q^2 + 40q^4 ) a_{1,0} c_{1,0}
			-3( 4 + 3k_0^2 - 64q^2 + 60q^4) c_{1,0}^2
		\Big],
 \nonumber
 \\
 &&
 	c_{2,0}^{\prime}(1)
 	=
 	\frac{1}{ 4(1  - q^2 )^2 }
 		\Big[
				8( -1 + k_0^2 + 4q^2 - 3q^4 ) c_{2,0}
			   	+8( -1 + q^2 )^2 a_{2,0}
				\nonumber
				\\
				&&
				\hspace{0.5cm}
				+ 4( -1 + q^2 )^2 a_{1,0}^2
				+( -8 + k_0^2 + 32q^2 - 24q^4 ) a_{1,0} c_{1,0}
				+( 4 + 3k_0^2 - 40q^2 + 36q^4 ) c_{1,0}^2
		\Big],
 \label{eq:regularity-X20}
\end{eqnarray}
where all functions are evaluated at $y=1$. The regularity also requires $b_{2,0}(1)=a_{2,0}(1)$.

By taking $a_{2,0}(1)$ and $c_{2,0}(1)$ as shooting parameters, one can easily integrate the coupled ODEs (\ref{eq:EOM-X20}) with the boundary conditions (\ref{eq:X20-asymptotics}) and (\ref{eq:regularity-X20}).

%%%%%%%%%%%%%%%%%%%%%%%%%%%%%%%%%%%%%%%%%%%%%%%%%%%%%%%%%%%%%%%%%%%%%%%%%%%%%%%
\subsubsection{Zero mode: $X_{0,1}$}
\label{sec:zero-mode}
%%%%%%%%%%%%%%%%%%%%%%%%%%%%%%%%%%%%%%%%%%%%%%%%%%%%%%%%%%%%%%%%%%%%%%%%%%%%%%%

The structure of EOMs for the $ X_{0,1} $ mode is different from those for inhomogeneous modes.
The starting point is to decouple $b_{0,1}$ and $c_{0,1}$ from $a_{0,1}$.
%%%-----------------
\footnote{
This procedure is peculiar to this charged system. 
For the neutral case ($q=0$), it is possible to decouple $b_{0,1}$ from $a_{0,1}$ and $c_{0,1}$, and then we can solve $b_{0,1}$.
Since such a equation for $b_{0,1}$ contains only derivatives of $b_{0,1}$, there is a shift symmetry and thereby we can integrate it without shooting~\cite{Gubser:2001ac}.
However, because of new terms coming from gauge fields, it is hard to obtain a single equation for $b_{0,1}$.  
}
%%%------------------
The resultant equations for $b_{0,1}(y)$ and $c_{0,1}(y)$ take the form of
\begin{eqnarray}
 &&
	b_{0,1}^{ \prime\prime }
	+
	\frac{1}{ y^2 ( -1 + y^2 )( 2q^4 - 5q^2y^2 + 3y^4) }
		\Big[
			y^3 [ 5q^4 - 3q^2 + 9y^4 + q^2(1-12y^2) ] b_{0,1}^{\prime}
 			\nonumber
 			\\
 			&&
 			\hspace{1.5cm}
		 	-3q^2y( q^2 - y^2 )( -2 + y^2 ) c_{1,0}^{\prime}
			-2q^2[ 2q^2 + 3y^2( -2 + y^2 ) ] b_{0,1}
			-6q^2( 2q^2 - 6y^2 + 4y^4 ) c_{0,1}
		\Big] 
	 \nonumber
 	\\
 	&&
 	\hspace{12cm}
	=
	S_{0,1}^{[1]} \left( X_{1,0}; k_0 \right),
 \nonumber
 \cr
 &&
	c_{0,1}^{\prime\prime}
	 +
	\frac{1}{ y^2( -1 + y^2 )( 2q^4 - 5q^2y^2 + 3y^4 ) }
		\Big[
			y \left[
				3y^4 ( -3 + 4y^2 ) + q^4( -4 + 5y^2 ) + q^2 ( 11y^2 - 15y^4 )
			  \right]c_{0,1}^{\prime}
			\nonumber\\
			&&
			\hspace{0.1cm}
			-y( q^2 - y^2 )^2 ( -2 + 3y^2 ) b_{0,1}^{\prime}
			+[ -4q^4 - 6y^6 + 2q^2y^2 ( 4 + y^2 ) ] b_{0,1}
			+2[ 6q^4 + 3y^6 - q^2y^2 ( 12 + y^2 ) ] c_{0,1}
		\Big]
	 \nonumber
	 \\
	 &&
 	 \hspace{12cm}
	=
	S_{0,1}^{[2]} \left( X_{1,0}; k_0 \right),
 \label{eq:EOM-X01}
\end{eqnarray}
where $S_{0,1}^{[i]}$ represents a source term, which is quadratic in $X_{1,0}$ and its first derivative.

Since the source terms decay exponentially in the asymptotic region, we can neglect them there. By the asymptotic expansion, we find $b_{0,1}(y)$ and $c_{0,1}(y)$ have power-law asymptotics, which are given in Eq.~(\ref{eq:asymptotics}).
Here, noting that the constants in Eq.~(\ref{eq:asymptotics}) are $O(\epsilon^2)$, we write them as $ B_{\infty} \simeq \epsilon^2 b_{\infty} $, $ C_{\infty} \simeq \epsilon^2 c_{\infty} $, $ C_{1} \simeq \epsilon^2 c_{1} $ (and $A_{\infty} \simeq \epsilon^2 a_{\infty}$). 
From the regularity at the horizon, the horizon boundary conditions are 
\begin{eqnarray}
&&
b_{0,1}^{\prime}|_{y=1}
	 =
			q^2 (6c_{0,1} -a_{1,0}^2 - 2b_{0,1} )
			+\frac{3}{4} \left( \frac{ k_0^2}{ 1-q^2 } + 8q^2 \right) a_{1,0} c_{1,0}
		 	-\frac{3}{4} \left( \frac{ k_0^2}{ 1-q^2 } - 12q^2  \right) c_{1,0}^2 ,
% \nonumber
 \\
 &&
	c_{0,1}^{\prime}|_{y=1}
		 =
			- 2 ( 1-3q^2 ) c_{0,1}		 
			+ (1-q^2) ( a_{1,0}^2 + 2b_{1,0}^2 )
			-  \frac{ 8 + k_0^2 - 32q^2 + 24q^4 }{ 4 ( 1-q^2 ) } a_{1,0} c_{1,0}
			\nonumber
			\\
&&
			\hspace{10cm}
			+  \frac{ 4 - 3k_0^2 - 40q^2 + 36q^4 }{ 4 ( 1-q^2 ) } c_{1,0}^2.
 \label{eq:X01-BC}
\end{eqnarray}
The equation for $a_{0,1}$ takes the form of
\begin{eqnarray}
&&
	a_{0,1}^{\prime\prime}
	 +
	 \frac{1}{ y^2( y^2-q^2 )( y^2-1) }
	 	\nonumber
	\\
&&
	\hspace{1cm}
	\times
		\Big[	
			y[ q^2( 2+y^2) -3y^4 ] a_{0,1}^{\prime}
			+ 3y( q^2 - y^2 ) c_{0,1}^{\prime}
		    +4q^2 (b_{0,1}-3c_{0,1})
		\Big]
	 =
	S_{0,1}^{[3]}(X_{1,0}; k_0).
 \label{eq:EOM-a01}
\end{eqnarray}
The asymptotic behavior is given by Eq.~(\ref{eq:asymptotics}) again. The regularity at the horizon requires
\begin{eqnarray}
&&
	a_{0,1}^{\prime}|_{y=1}
	 =
	-\frac{1}{ 12( 1 - q^2 )^2 }
		\Big[
			2( 6 + k_0^2 - 16q^2 + 10q^4 ) a_{1,0}^2
			-8(1-q^2)
			[
				( 5q^2-3) b_{0,1}
				\nonumber \\
				&&
				\hspace{1cm}
				+ 3( 1 - 5q^2 ) c_{0,1}
				+3(  k_0^2 -8 + 48q^2 - 40q^4 ) a_{1,0} c_{1,0}
				+3( 4 - 3k_0^2 - 64q^2 + 60q^4 ) c_{1,0}^2
			]	
		\Big],
 \label{eq:BC-a01}
\end{eqnarray}
Since  $a_{0,1}$ is decoupled from $b_{0,1}$ and $c_{0,1}$, all the values in the right-hand side (RHS) will be known after the integration of $b_{0,1}$ and $c_{0,1}$.
Furthermore, we notice that the above equation (\ref{eq:EOM-a01}) and the boundary condition (\ref{eq:BC-a01}) contain only the first and second derivatives of $a_{0,1}(y)$. This means that $a_{0,1}$ has a shift symmetry, $ a_{0,1} \to a_{0,1}  +  \mathrm{constant}$, corresponding to the gauge degree of freedom of redefining a time coordinate.
Therefore, after integrating (\ref{eq:EOM-a01}) with an arbitrary value of $a_{0,1}(1)$, one can use the shift symmetry in order to realize $ a_{0,1} \to 0 $ at the asymptotic region. No shooting is necessary.

Finally, we discuss the integration of Eq.~(\ref{eq:EOM-X01}) for $b_{0,1} $ and $c_{0,1} $. It seems that we have two constants to be fixed by asymptotic flatness, the horizon values of $b_{0,1} $ and $c_{0,1} $.
However, we face the known puzzle  (for the uncharged case)~\cite{Gubser:2001ac,Wiseman:2002zc,Kol:2006vu}: any choice of $c_{0,1}(1)$ results in a regular solution satisfying asymptotic flatness. 
The physical quantities depend on the choice of $c_{0,1}(1)$, and thereby the physical quantities cannot be fixed at this order. 
This apparent contradiction can be understood by the fact that the general conformal ansatz (\ref{eq:conformal}) has the freedom to ``superpose'' arbitrary $O(\epsilon^2)$ change in the mass on the top of the change that the non-uniformity induces~\cite{Gubser:2001ac,Wiseman:2002zc}. 
The same argument holds for the present charged case: the equation (\ref{eq:EOM-X01}) with the source term set to zero, i.e., homogeneous equation, has a solution that corresponds to the $s$-wave perturbation of a $d$-dimensional charged black hole. 
The increase of mass and charge of the black hole is expressed as a solution to the homogeneous equations with the asymptotics (\ref{eq:asymptotics}). Any multiple of such a solution to the homogeneous equation can be added to the solutions to the inhomogeneous equations. This is the origin of the ambiguity.

Indeed, the horizon values of homogeneous solutions can be related to the change of physical quantities as follows. 
Consider a uniform black string given by setting $X(r,z)$ to $X(r)$ ($X=a,b,c$) in Eq.~(\ref{eq:string}). Introducing a conformal radial coordinate 
\begin{eqnarray}
	\rho (r;r_{\pm})
	 =
	\int^{r}  \frac{ dr' }{ \surd f_{+}(r'; r_{+}) f_{-}^{2}(r'; r_{-}) } ,
 \label{eq:conformal-tr}
\end{eqnarray}
the metric can be written as
\begin{eqnarray}
	ds_{d+1}^2 
	 &=&
	- e^{2a} f_+  dt^2 
	+  e^{2b} f_-  \left( d\rho^2 + dz^2 \right)
	+  e^{2c} r^2  d\Omega_{d-2}^2.
 \label{eq:conformal-form}
\end{eqnarray}
In Eq.~(\ref{eq:conformal-form}), $r$ should be regarded as a function of $\rho$, which we denote by $r=r(\rho, r_{\pm})$.
The integration constant of Eq.~(\ref{eq:conformal-tr}) can be fixed so that $\rho = 0 \leftrightarrow r=r_{+}$.
Because of the three functional degrees of freedom, there is a solution that changes mass and charge of the background black string.
Such solutions with new parameters $r_\pm'$ are given by the following homogeneous shifts of metric functions:
\begin{eqnarray}
 &&
	X( r(\rho;r_{\pm}) ) \to X( r(\rho;r_{\pm}) ) + \Delta_X( \rho; r_{\pm}, r'_{\pm} ), 
 \nonumber
 \\
 &&
	\Delta_a
%	(\rho; r_{\pm}, r'_{\pm})
	 = 
	\frac{1}{2}
		\ln
			\left[
				\frac{ f_{+} ( r( \rho; r'_{\pm}); r'_{+} ) }
					 { f_{+} ( r( \rho; r_{\pm}); r_{+}   ) }
			\right], \;\;\;\;
% \nonumber
% \\
% &&
	\Delta_b
%	(\rho; r_{\pm}, r'_{\pm})
	 =
	\frac{1}{2}
		\ln
			\left[
				\frac{ f_{-} ( r( \rho; r'_{\pm}); r'_{-} ) }
					 { f_{-} ( r( \rho; r_{\pm}); r_{-}   ) }
			\right], \;\;\;\; 
% \nonumber
% \\
% &&
	\Delta_c
%(\rho; r_{\pm}, r'_{\pm})
	 =
		\ln
			\left[
				\frac{ r(\rho;r'_{\pm}) }{ r(\rho;r_{\pm}) }
			\right],
 \label{eq:shift}
\end{eqnarray} 
where we write $f_{\pm}(r)$ as $f_{\pm}(r;r_{\pm})$ in order to show explicitly its dependence on the parameter $r_{\pm}$.  
Corresponding to these shifts, we must also change the charge that appears in the RHS of the Einstein equation through the field strength. 
With all these transformation, the resultant solution is the charged uniform black strings with new parameter $r_{\pm}'$. 
It may be useful to derive the relation between the horizon values of the shifts and parameters $r'_{\pm}$. Taking the limit $\rho \to 0$ in (\ref{eq:shift}), we have
\begin{eqnarray}
	\lim_{\rho\to 0} \Delta_a
	 =
	-\ln \left( \frac{r'_{+}}{r_{+}} \right)
	 +
	\ln \left( \frac{ 1-q^{\prime d-3} }{ 1-q^{d-3} } \right),
\quad
	\lim_{\rho\to 0} \Delta_b
	 =
	\frac{1}{2} \ln \left( \frac{ 1-q^{\prime d-3} }{ 1-q^{d-3} } \right),
\quad
	\lim_{\rho\to 0} \Delta_{c}
	 =
	\ln \left( \frac{ r'_{+} }{ r_{+} } \right),
 \label{eq:shift-horizon}
\end{eqnarray}
where $q' \equiv r'_-/r'_+$ and we have used $\lim_{\rho\to 0}r(\rho; r_{\pm}) = r_+$.

At the practical level of calculations (perturbative expansions, etc.), it is easy to change $c_{0,1}(1)$, but is cumbersome to perform the necessary shift of the parameter $r_\pm$ which comes in through the field strength. 
If we forget the required shift of the parameter $r_\pm$ in the field strength and change only $c_{0,1}(1)$ in our approach, the new background charge is automatically fixed and we have no freedom to perform an arbitrary shift of $b_{0,1}(1)$. 
In fact, as Eq.(\ref{eq:shift-horizon}) predicts, our numerical solutions confirm that 
\begin{eqnarray}
   (a_{0,1} -2b_{0,1} + c_{0,1})_{y=1}
\end{eqnarray}
is invariant under the shift of $c_{0,1}(1)$, in which $b_{0,1}(1)$ is automatically determined as a result of the shooting procedure.
In the present paper, we want to construct the charged non-uniform string for a fixed charge and compare its physical quantities at the fixed charge. 
For this reason and to perform a clear analysis, we set $c_{0,1}(1) = 0 $ and solve Eq.~(\ref{eq:EOM-X01}) by taking $ b_{0,1}(1) $ as the shooting parameter. 
The shooting parameter resulting in a regular solution and the asymptotic constants for the neutral case are given in Table~\ref{table:boundary-values}. 
As a consistency check, we will show later that the final result does not depend on the choice of $c_{0,1}(1)$
(see Fig.~\ref{fig:6D-sigmarho}).

%%%%%%%%%%%%%%%%%%%%%%%%%%%%%%%%%%%%%%%%%%%%%%%%%%%%%%%%%%%%%%%%%%%%%%%%%%%%%%%
\begin{table}[tbp]
\begin{center}
			\setlength{\tabcolsep}{9pt}
			\begin{tabular}{  c | c | c | c | c | c | c | c | c | c | c | c  }
 \hline\hline
 	$D$			& $a_{1,0}$ 	& $k_0$ 			& $a_{2,0}$ 		&
	$c_{2,0}$ 	& $a_{0,1}$ 	&	 $b_{0,1}$ 		& $a_\infty$ 		&
	$b_\infty$ 	& $c_\infty$ 	& $a_{1,1}$ 		& $k_1$  
 \\
 \hline 
	6			&  -0.78		&  	1.27			&  	0.64			&
	-0.87 		& 	0.63		&	0.91  			&  	-0.93			&
	0.65	 	&  	-0.37		&  	-0.38			&   1.02
 \\
	10			&  -1.54		&  	2.30			&  	2.48			&
	-1.53 		& 	0.66		&	1.46  			&  	-4.06			&
	1.67	 	&  	0.14		&  	-0.44			&   0.12
 \\
	14			&  -2.15		&  	3.03			&  	5.16			&
	-2.14 		& 	0.43		&	1.90  			&  	-10.0			&
	2.69	 	&  	0.51		&  	1.77			&   -6.13
 \\   
 \hline
 \hline
		\end{tabular}
\end{center}
		\caption{
Shooting parameters (horizon values of metric perturbations and critical wavenumbers) and asymptotic quantities for the regular normalizable solutions with $Q=0$. 
See section~\ref{sec:perturbation} for their definitions. We can use these results as initial-guess values for charged cases, $Q \neq 0$.
		\label{table:boundary-values} }
\end{table}
%%%%%%%%%%%%%%%%%%%%%%%%%%%%%%%%%%%%%%%%%%%%%%%%%%%%%%%%%%%%%%%%%%%%%%%%%%%%%%%

%%%%%%%%%%%%%%%%%%%%%%%%%%%%%%%%%%%%%%%%%%%%%%%%%%%%%%%%%%%%%%%%%%%%%%%%%%%%%%%
\subsection{Third-order perturbation: $X_{1,1}$}
%%%%%%%%%%%%%%%%%%%%%%%%%%%%%%%%%%%%%%%%%%%%%%%%%%%%%%%%%%%%%%%%%%%%%%%%%%%%%%%

In the present approach, it is necessary to know the correction to the GL critical mode, $k_1$, which appears at third order $O(\epsilon^3)$.
At this order, we have two independent modes, $X_{1,1}$ and $X_{3,0}$. 
However, since the equations for the latter mode $X_{3,0}$ do not contain $k_1$, it is sufficient for our purpose to solve $X_{1,1}$.

The structure of EOMs and boundary conditions for $X_{1,1}$ are similar to those for the $X_{1,0}$ and $X_{2,0}$ modes. By decoupling $b_{1,1}(y)$ from $a_{1,1}(y)$ and $c_{1,1}(y)$, we obtain EOMs for $a_{1,1}(y)$ and $c_{1,1}(y)$, taking the form of
\begin{eqnarray}
 &&
 	\hat{L}_{k_0}^{[1]} a_{1,1}
 	 +
 	\hat{P}^{[1]} c_{1,1}
 	 =
 	S_{1,1}^{[1]} ( X_{1,0}, X_{2,0}, X_{0,1}; k_0, k_1 ),
 \nonumber
 \\
 &&
 	\hat{L}_{k_0}^{[2]} c_{1,1}
 	 +
 	\hat{P}^{[2]}a_{1,1}
 	 =
 	S_{1,1}^{[2]} ( X_{1,0}, X_{2,0}, X_{0,1}; k_0, k_1 ).
 \label{eq:X11-EOMs}
\end{eqnarray}
The function $b_{1,1}(y)$ can be written algebraically in terms of $a_{1,1}(y)$ and $c_{1,1}(y)$. 
By solving (\ref{eq:X11-EOMs}) in the asymptotic region, we obtain asymptotic behaviors,
\begin{eqnarray}
	b_{1,1}(y),\; 
	c_{1,1}(y)
	 \sim
	e^{-k_0 y}.
 \label{eq:bc11asym}
\end{eqnarray}
From the regularity at the horizon,
$a'_{1,1}(1)$ and $c'_{1,1}(1)$ can be written in terms of $a_{1,1}(1)$, $c_{1,1}(1)$ and the values and first derivatives of $ X_{1,0} $ at the horizon.

Which parameters should be determined by shooting at this order? 
According to~\cite{Gubser:2001ac}, the different choice of $k_1$ corresponds to a different ``scheme'' (gauge).  
We adopt the ``standard'' scheme in which we set 
$ c_{1,p}(1) = 0 $ for $ p>0 $, and permit $ k_1 $ to vary.\footnote{
It is also possible to set $ k_1 = 0 $ in addition to $ c_{1,p} = 0 $ ($ p>0 $).
However, this requires a technically difficult procedure. 
In this case, $c_{0,1}(1)$ at the second order should be dealt with as a shooting parameter so that the third-order perturbations converge at all. 
It means that the second- and third-order perturbations have to be solved at the same time.
}
Setting $c_{1,1}(1)$ equal to zero, we determine $a_{1,1}(1)$ and $k_1$ by requiring normalizable modes.
Namely, we integrate Eq.~(\ref{eq:X11-EOMs}) with the horizon regularity condition and the asymptotic damping (\ref{eq:bc11asym}) by taking $a_{1,1}(1)$ and $k_1$ as the shooting parameters, as we do at the first order.

%%%%%%%%%%%%%%%%%%%%%%%%%%%%%%%%%%%%%%%%%%%%%%%%%%%%%%%%%%%%%%%%%%%%%%%%%%%%%%%
\section{Physical properties of inhomogeneous charged strings}
\label{sec:physical properties}
%%%%%%%%%%%%%%%%%%%%%%%%%%%%%%%%%%%%%%%%%%%%%%%%%%%%%%%%%%%%%%%%%%%%%%%%%%%%%%%

%%%%%%%%%%%%%%%%%%%%%%%%%%%%%%%%%%%%%%%%%%%%%%%%%%%%%%%%%%%%%%%%%%%%%%%%%%%%%%%
\subsection{Changes of physical quantities}
\label{sec:thermo1}
%%%%%%%%%%%%%%%%%%%%%%%%%%%%%%%%%%%%%%%%%%%%%%%%%%%%%%%%%%%%%%%%%%%%%%%%%%%%%%%

\begin{figure}[t]
	\begin{center}
	\includegraphics[width=8cm]{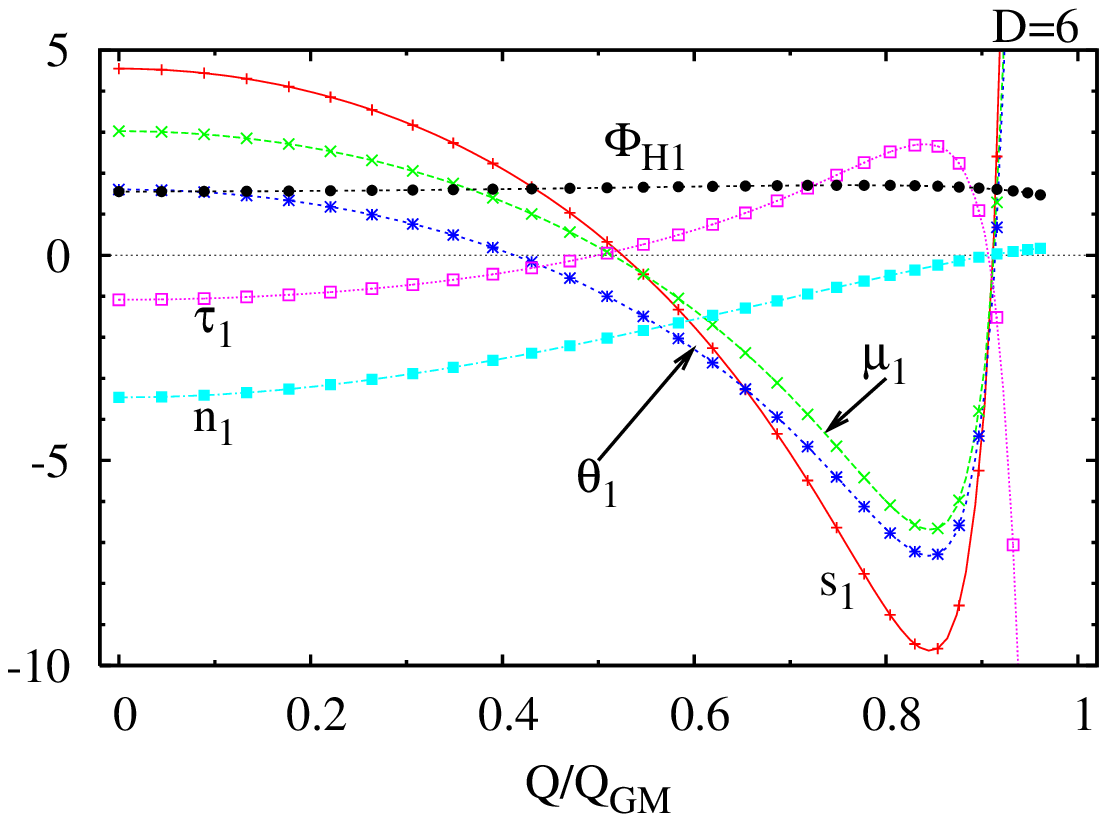}
	\includegraphics[width=8cm]{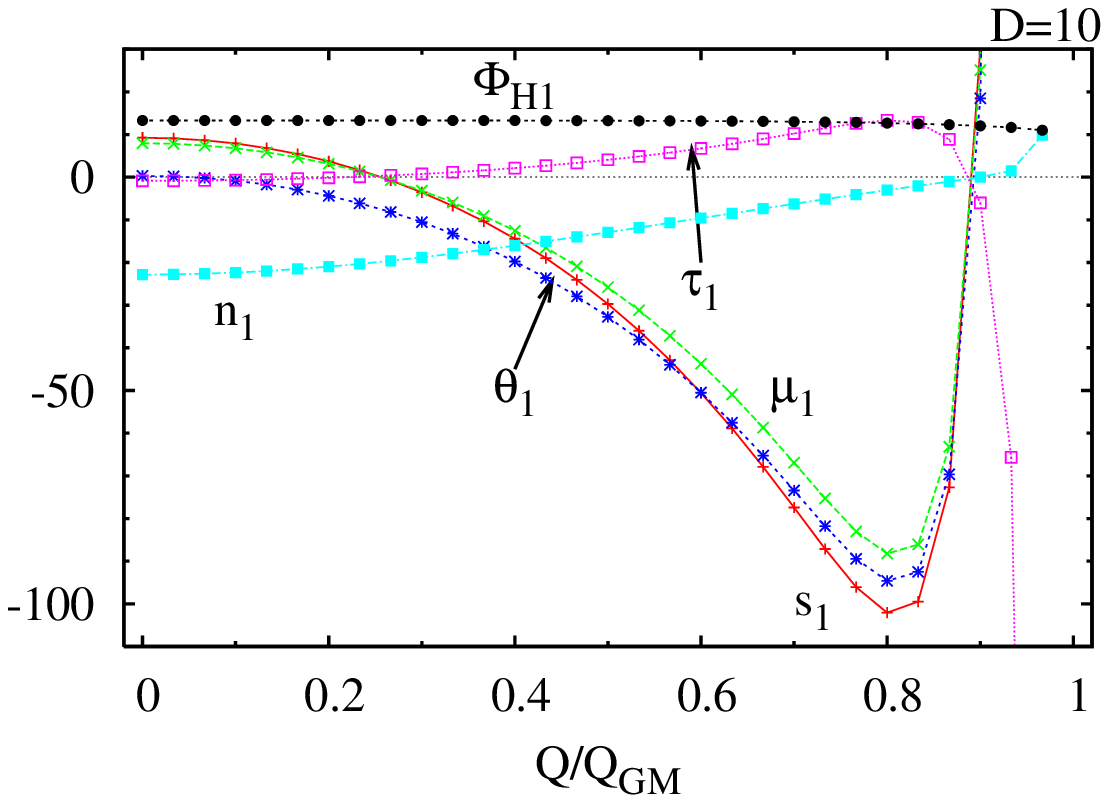}
	\includegraphics[width=8cm]{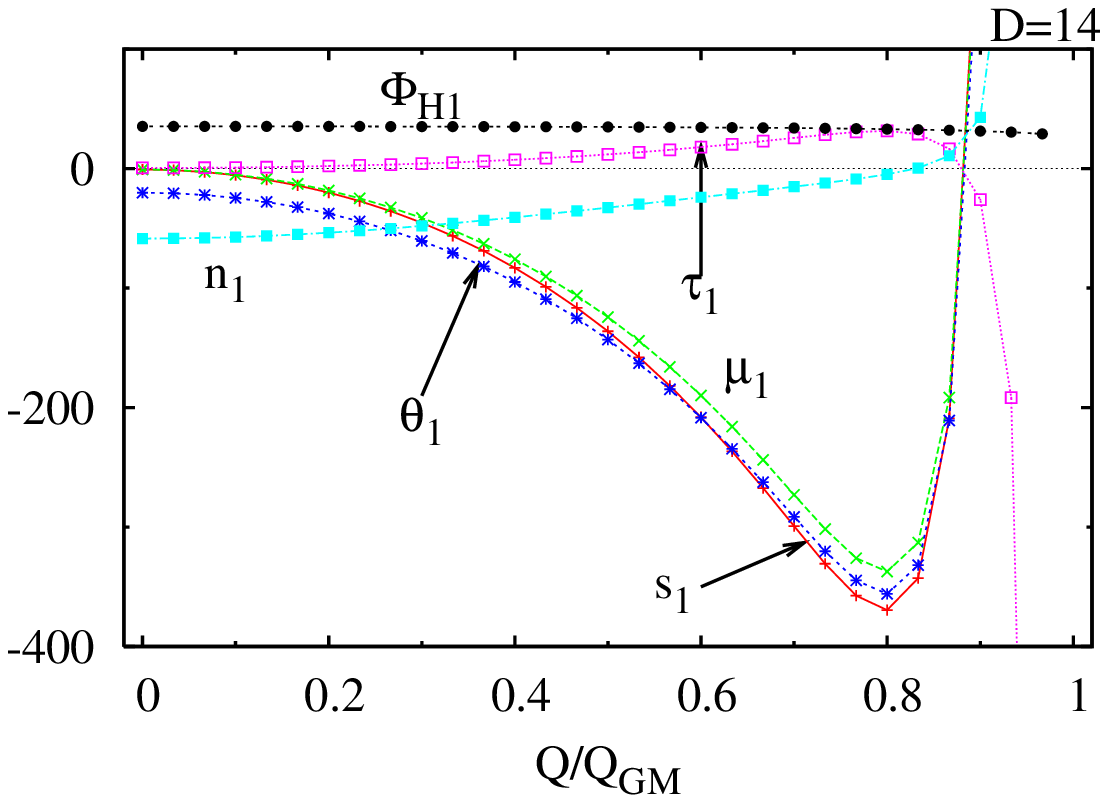}	
		\caption{
The differences of physical quantities (\ref{eq:thermo-diff}) and (\ref{eq:invariants}) between the non-uniform and the critical uniform black strings for each value of background charge $Q$ for $D=6, 10, 14$.
For $D=6$ and $10$ ($\leq 13$) one can see that all quantities except $\Phi_{H1}$ change their signs.
To obtain the results, we have adopted $c_{0,1}(1)=0$. 
\label{fig:6D-thermo s1 mu1 etc}
}
	\end{center}
\end{figure}

With the formulae prepared in Sec.~\ref{sec:Physical quantities} and perturbative quantities calculated in Secs.~\ref{sec:static-mode} and \ref{sec:perturbation}, the increase of thermodynamical quantities due to the non-uniform deformation can be calculated.
Because the asymptotic size of the transverse circle is not fixed ($k_1 \neq 0$), we introduce variables that are invariant under rigid rescalings of the entire solution.
Such invariant quantities can be obtained by multiplying $K$ by suitable powers.
The relative increase of the variables is given by
\begin{eqnarray}
 &&
	\frac{ \delta \mu }{ \mu }
	 \equiv
	\frac{ \delta M }{ M } + ( d - 3 ) \frac{ \delta K }{ K }
	 =
	\mu_{1} \epsilon^{2} 
	 +
	O(\epsilon^{4}),
 \cr
 &&
	\frac{ \delta s }{ s }
	 \equiv
	\frac{ \delta S }{ S } + (d-2) \frac{ \delta K }{ K }
	 =
	s_{1} \epsilon^{2} 
	 +
	O(\epsilon^{4}),
\cr
 &&
 	\frac{ \delta \tau }{ \tau }
	 \equiv
	\frac{ \delta T }{ T } - \frac{ \delta K }{ K }
  	 =
	\tau_{1} \epsilon^{2}
	 +
	O( \epsilon^{4} ),
 \cr
 &&
 	\frac{ \delta \vartheta }{ \vartheta }
	 \equiv
	\frac{ \delta Q }{ Q } + (d-3) \frac{ \delta K }{ K }
	 =
	\vartheta_{1} \epsilon^{2}
	 +
	O( \epsilon^{4} ),
 \label{eq:thermo-diff}
\end{eqnarray}
where the second-order coefficients (for $d \geq 5$) are given by
\begin{eqnarray}
&&
	\mu_{1}=
%		\begin{cases}
%			\displaystyle
%				\frac{2(3b_{\infty}-2c_{\infty})}{2+q}  +  \frac{k_1}{k_0}
%				 &
%				\mbox{for $d=4$},
%				 \\
%			\displaystyle
				\frac{ 2( 2d - 5 ) b_{\infty} + 2(d-2)(d-4) c_{\infty} }
				     {d-2+q^{d-3}}
				+ (d-3) \frac{ k_{1} }{ k_0 },
%	  		     &
%	  		    \mbox{ for $d\geq 5$ },
% \end{cases}
 \cr
 &&
 	s_{1}
 	 =
 	b_{0,1}  + (d-2) c_{0,1} + 
			\frac{    \left[ b_{1,0}  + (d-2) c_{1,0} 
	\right ]^2 }{4} + (d-2) \frac{ k_{1} }{ k_{0} },
 \cr
 &&
	\tau_{1}
	 =
	 a_{0,1}- b_{0,1} - \frac{ k_{1} }{ k_{0} },
 \cr
 &&
 	\vartheta_{1} = (d-3) \frac{ k_1 }{ k_0 }.
 \label{eq:invariants}
\end{eqnarray}
The potential and relative tension are invariant by themselves, and their relative increase is given by
\begin{eqnarray}
	&&
	\frac{ \delta \Phi_H }{ \Phi_H }
	 =
	\Phi_{H1} \epsilon^2
	 +
	O(\epsilon^4), 
	\;\;\;\;\;
	\frac{\delta n}{n}
	 = 
	n_1 \epsilon^2
	 +
	O(\epsilon^4),
	\nonumber
	\\
	&&
	\Phi_{H1}
	=
	(d-3) \int_{1}^{\infty} \frac{ dy }{ y^{d-2} }
		\left\{
			a_{0,1} + 2 b_{0,1} -(d-2)c_{0,1}
			+
			\frac{1}{4}
				\left[
					a_{1,0}+2b_{1,0}-(d-2)c_{1,0}
				\right]^2
		\right\},
	\nonumber
	\\
	&&
	n_1
	=
	- \frac{ 2 (d-1)(d-3) [ (1+2q^{d-3})b_{\infty} + (d-4) q^{d-3} c_{\infty} ]}
		   { ( d - 2 + q^{d-3} )[ 1 + (d-2)q^{d-3} ] }.
	\label{eq:invariants2}
\end{eqnarray}
The charge dependence of these quantities are depicted in Fig.~\ref{fig:6D-thermo s1 mu1 etc} for $D=6,10,14$.
For $D=6$ and $10$, one can observe that the change of mass $\mu_1$, being positive initially at $Q=0$, decreases as the background charge increases. 
Then, $\mu_1$ becomes negative, e.g., at $Q \simeq 0.5 Q_{ \mathrm{GM} }$ for $D=6$. 
The increase of entropy and charge, $s_1$ and $\vartheta_1$, behaves in a similar way. The temperature $\tau_1$ also behaves in a similar way but in the opposite sign; the non-uniform black string, being ``cooler'' than the critical solution initially at $Q=0$, becomes ``hotter'' as $Q$ increases.
Some remarkable properties are that
$\Phi_H$ does not change the sign and $n_1$ changes its sign near the GM point where $s_1$, $\mu_1$, $\theta_1$, $\tau_1$ also cross the zero axis. 
The positive value of $n_1$ has an interesting physical meaning. 
The uniform black string is not necessarily the state that has the largest tension (for a fixed background charge).
This is not impossible for uncharged black stings.

In Fig.~\ref{fig:6D-thermo s1 mu1 etc}, we have shown the results up to $Q/Q_{\mathrm{GM}} \simeq 0.95$. 
The physical quantities in Eq.~(\ref{eq:invariants}) contain the term of $k_1/k_0$, which diverges in the limit of $Q \to Q_{\mathrm{GM}}$, and thereby numerical accuracy becomes relatively worse near the GM point.
To confirm the consistency of analysis and the accuracy of numerical integration, we use the Smarr formula as an independent check.

According to Ref.~\cite{Townsend:2001rg, Harmark:2004ch}, we find the following result: 
\begin{eqnarray}
(D-3) {\mathcal{E}} - (D-2) T \mathcal{S}  - \mathbb{T}
 +  \mathcal{E}^{(mat)}  =0,
\label{eq:smarr original form}
\end{eqnarray}
where 
$ {\mathcal{E}}$ and $\mathcal{S}$
are the mass and entropy density along the $z$-direction, respectively, and $\mathbb{T}$ is the tension.
$\mathcal{E}^{(mat)}$ is the mass density for matter fields evaluated on the horizon. It is given by the surface integral on the horizon at $z$, 
\begin{eqnarray}
\mathcal{E}^{(mat)} (z)
 &=& \frac{1}{(p+2)!}
   \oint _H dS_{MNP_1 \cdots P_p} \mathcal{F}^{MNP_1 \cdots P_p} \phi_H(r_+,z),
\end{eqnarray}
where $\phi_H(r_+,z)$ is the density of electric potential at $z$. 
Taking the dual of Eq.~(\ref{eq:general-F}) and substituting it into the above formula, 
the mass density for the matter field is found to be
$\mathcal{E}^{(mat)}=  (Q/L)\phi_H $. 
Integrating (\ref{eq:smarr original form}) over the $z$ direction yields the Smarr formula which holds even for the non-uniform black strings:
\begin{eqnarray}
(D-3) M - (D-2) TS  - L \mathbb{T}
 +  Q \Phi_H =0.
\end{eqnarray}
This formula can be also rewritten in terms of relative tension $n$ as 
\begin{eqnarray}
	\frac{TS}{M} + \frac{d-3}{d-1} \Phi_H \left(  \frac{Q}{M} \right)
  	 =
  	\frac{ d-2-n }{ d-1 }.
 \label{eq:smarr}
\end{eqnarray}

Now we expand all physical quantities as power series of parameter $\epsilon$:
\begin{eqnarray}
	Y
	 =
	\sum_{ p=0 }^{ \infty } Y_{p} \epsilon^{2p},
	 \quad
	( Y = M, S,~\mathrm{etc}.),
 \label{eq:expansion of MSTQF}
\end{eqnarray}
Substituting the expansion into Eq.~(\ref{eq:smarr}), we obtain the following relation at $O(\epsilon^2)$,
\begin{eqnarray}
  \delta S_1 + \delta T_1 - \delta M_1
   +
  \frac{n_0}{d-2 -n_0} \delta n_1
   +
  \frac{ (d-3)Q_0 \Phi_{H0} }{ (d-1) S_0 T_0 }
  	\left( 
   		\frac{ n_0 }{ d-2-n_0 } \delta n_1
   		 -
   		\delta M_1 + \delta Q_1 + \delta \Phi_{H1}
    \right)
     =
    0,
\label{eq:leading-smarr}
\end{eqnarray}
where we abbreviate $\delta Y_1 \equiv Y_1/Y_0$.
We can replace the quantity $\delta Y_1$ by the scale invariant quantity, e.g., $\delta M_1 \to \mu_1$, in Eq.~(\ref{eq:leading-smarr}).
Therefore, Eq.~(\ref{eq:leading-smarr}) is a constraint equation that the physical quantities must obey. 
Substituting our solutions into the Smarr formula, we have confirmed that the Smarr formula holds within a relative error of only a few percent.

%%%%%%%%%%%%%%%%%%%%%%%%%%%%%%%%%%%%%%%%%%%%%%%%%%%%%%%%%%%%%%%%%%%%%%%%%%%%%%%
%%%%%%%%%%%%%%%%%%%%%%%%%%%%%%%%%%%%%%%%%%%%%%%%%%%%%%%%%%%%%%%%%%%%%%%%%%%%%%%

%%%%%%%%%%%%%%%%%%%%%%%%%%%%%%%%%%%%%%%%%%%%%%%%%%%%%%%%%%%%%%%%%%%%%%%%%%%%%%%
%%\begin{figure}[h]
%%	\begin{center}
%%	\includegraphics[width=8cm]{figs/10D-thermo-cp1.ps}
%%	\includegraphics[width=8cm]{figs/10D-thermo-cm1.ps}
%%\caption{
%%The plot of physical quantities (\ref{eq:invariants}) for $D=10$ with $c_{0,1}= \pm1$ case. 
%%\label{fg:6D-thermo 2 c01=+-1}
%% }
%%	\end{center}
%% \end{figure}
%%%%%%%%%%%%%%%%%%%%%%%%%%%%%%%%%%%%%%%%%%%%%%%%%%%%%%%%%%%%%%%%%%%%%%%%%%%%%%%

%%%%%%%%%%%%%%%%%%%%%%%%%%%%%%%%%%%%%%%%%%%%%%%%%%%%%%%%%%%%%%%%%%%%%%%%%%%%%%%
\subsection{Thermodynamical stability in microcanonical and canonical ensembles}
\label{sec:sigmarho}
%%%%%%%%%%%%%%%%%%%%%%%%%%%%%%%%%%%%%%%%%%%%%%%%%%%%%%%%%%%%%%%%%%%%%%%%%%%%%%%

%%%%%%%%%%%%%%%%%%%%%%%%%%%%%%%%%%%%%%%%%%%%%%%%%%%%%%%%%%%%%%%%%%%%%%%%%%%%%%%
\begin{figure}[t]
	\begin{center}
	\includegraphics[width=8cm]{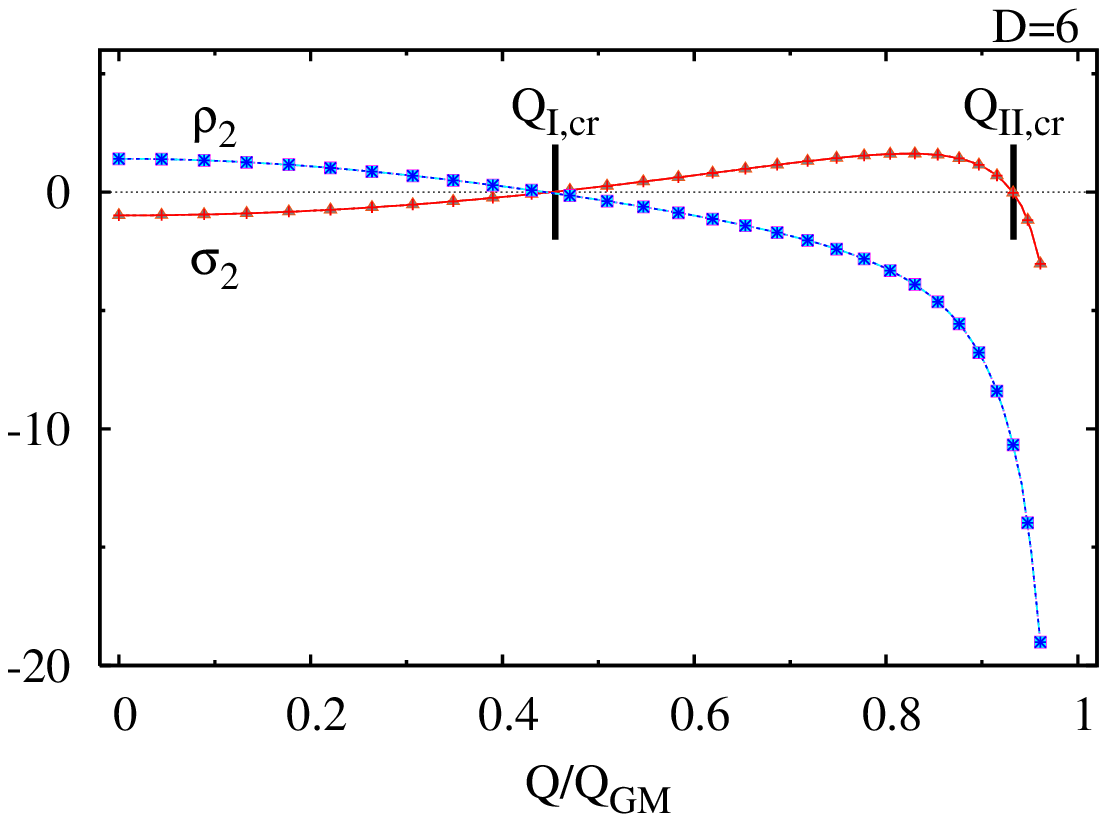}
	\includegraphics[width=8cm]{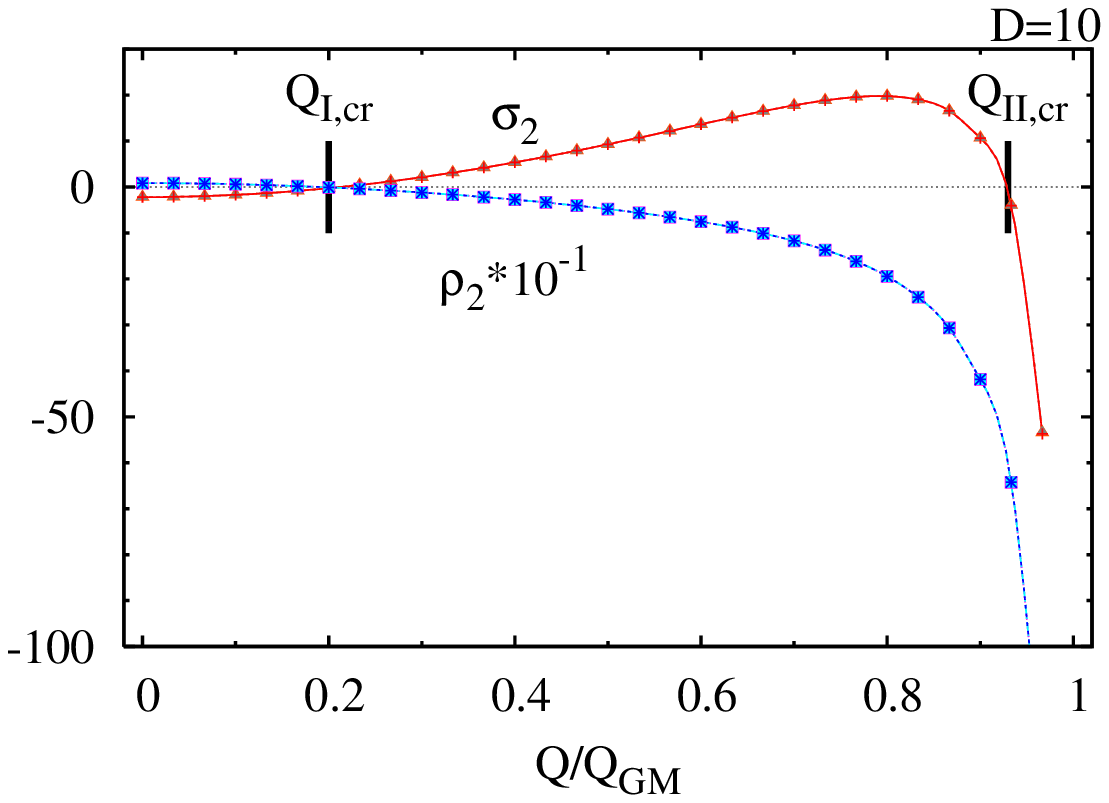}
	\includegraphics[width=8cm]{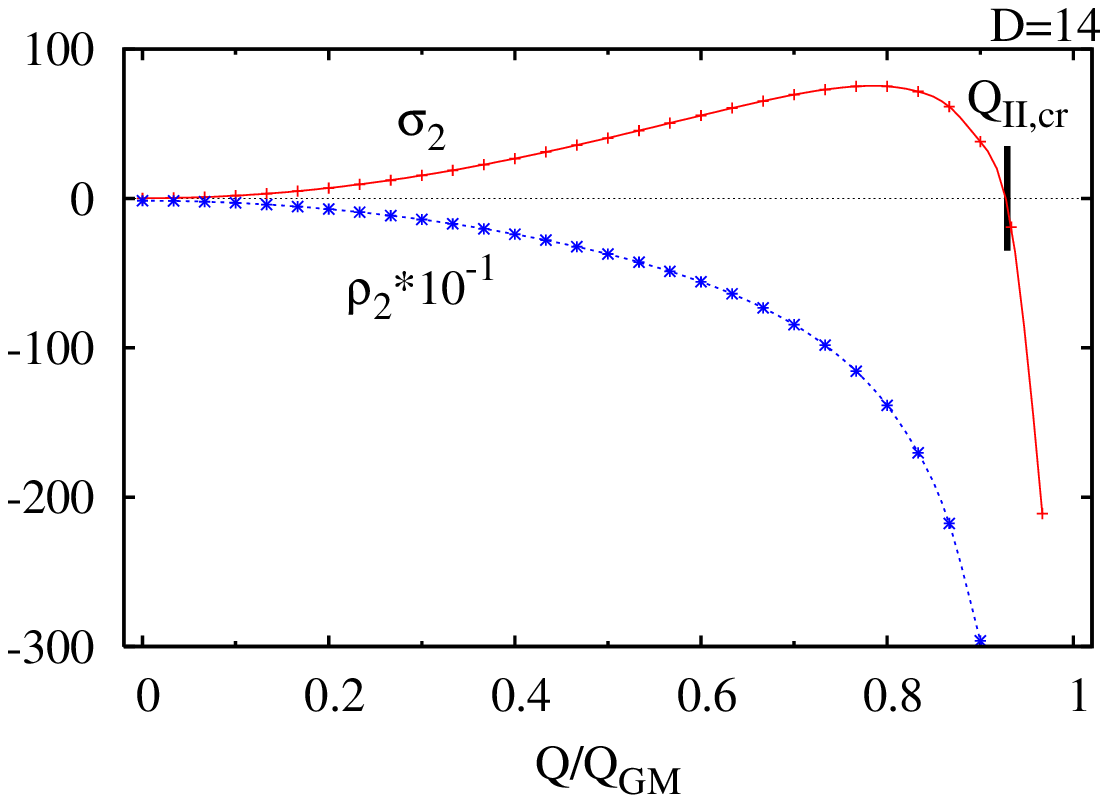}
\caption{
The difference of entropy $\sigma_2$ (free energy $\rho_2$) between the non-uniform and uniform black strings of the same mass (temperature) and charge for $D=6,10,14$. 
$ Q_{\mathrm{I,cr}} $ and $ Q_{\mathrm{II,cr}} $ with the vertical lines show the locations of the critical charges.
In each figure, we plot three results obtained by setting $c_{0,1}(1)=0, +1, -1$, although they cannot be distinguished from each other.
For $D\le 13$, the non-uniform string with $Q_{\mathrm{I,cr}}<Q<Q_{\mathrm{II,cr}}$ has larger entropy than the uniform string. 
For $D\ge 14$, the free energy difference is never positive, and the first critical charge does not exist. 
\label{fig:6D-sigmarho}
}
	\end{center}
\end{figure}
%%%%%%%%%%%%%%%%%%%%%%%%%%%%%%%%%%%%%%%%%%%%%%%%%%%%%%%%%%%%%%%%%%%%%%%%%%%%%%%

We are ready to study the stability of the inhomogeneous charged black strings.
First, let us focus on the entropy comparison in a microcanonical ensemble, namely, the entropy difference between the non-uniform black string and a corresponding uniform black string of the same mass and charge. Such a difference is given by
\begin{eqnarray}
 &&
 	\frac{ S_{ \mathrm{NU} } - S_{ \mathrm{U} } }{ S_{\mathrm{NU} } }
	 =
	\sigma_1 \epsilon^2
	 +
	\sigma_2 \epsilon^4
	 +
	O( \epsilon^6 ),
 \nonumber\\
 &&
 	\sigma_1
 	 =
	s_1
	 -
	\frac{ d - 2 + \Phi_{H0}^2  }{ ( d - 3 )( 1 - \Phi_{H0}^2  ) } \mu_1
	 +
	\frac{ ( d - 1 )\Phi_{H0}^2  }{ ( d - 3 )( 1 - \Phi_{H0}^2  ) } \vartheta_1,
 \nonumber \\
 &&
	 \sigma_2
	  =
	 \frac{ (d-1) \Phi_{H0}^2  ~ \vartheta_1 }{ d - 2 - \Phi_{H0}^2  }  
 		\left\{ 
   			\frac{ d-2 }{ 2(d-3) } 
   				\left[ 
   		 \vartheta_1 - \frac{ d-2-\Phi_{H0}^2}{(d-2)(1-\Phi_{H0}^2)} \Phi_{H1}
   				\right] - s_1
    	\right\}
\nonumber
\\
&&
\hspace{9cm}
	 -
  	\frac{s_1}{2}
    	\left[ \frac{1-(d-2)\Phi_{H0}^2 }{d-2 -\Phi_{H0}^2 } s_1
  			 + 
  			\tau_1
		\right]. 
	\label{eq:sigma2}
\end{eqnarray} 
These equations can be derived by starting from the frame with $\delta K=0$ and re-expressing the results in terms of the invariant quantities (\ref{eq:invariants}). (See Appendix \ref{appendix:entropy-diff} for the details of calculation.)
The first law of thermodynamics requires $\sigma_1=0$. 
The vanishing of $\sigma_1$ indeed holds with permissible numerical errors (a few percent relative errors) except near the GM point, e.g., $Q \gtrsim 0.95  Q_\mathrm{GM}$. 
Thus, the difference of entropy appears at $O(\epsilon^4)$.
Note that we have used $\sigma_1=0$ to simplify the expression of $\sigma_2$.
If $\sigma_2$ is positive, it means that the non-uniform state has larger entropy than the uniform state and the non-uniform state is more stable. 
If $\sigma_2$ is negative, the uniform state is more favorable. 
For the uncharged case, the latter case is realized for $D\le 13$, and the former case is for $D \ge 14$.

The dependence of $ \sigma_2 $ on the background charge $Q$ is depicted in Fig.~\ref{fig:6D-sigmarho} for $D=6,10,14$. 
As expected from the sign change of $ \mu_1 $, observed in the previous section, the sign of $ \sigma_2 $ changes depending on the background charge.
For $D\le 13$,  $ \sigma_2 $, being negative initially at $ Q=0 $, increases as the charge increases, and it becomes positive at some critical charge $Q=Q_{{\mathrm{I,cr}}}$. 
However, the increase is not monotonic, rather, $ \sigma_2 $ has a peak. 
Increasing the charge furthermore toward the GM point,  $\sigma_2$ falls off and becomes negative at a second critical charge $Q=Q_{\mathrm{II,cr}}$.
The numerical value of the second critical charge corresponds to the zero-crossing point of $n_1$ (or the second zero crossing of $s_1$, $\mu_1$, etc.). 
This behavior near the GM point is general. 
For instance, $ \sigma_2 $ for $D=10$ shows similar behavior. 
A quantitative difference is that the first critical charges are achieved by smaller charges, while the second critical charge does not change so much. 
For $D \ge 14$, since $\sigma_2$ is initially positive at $Q=0$, the first critical charge that we observed for $D\le 13$ is absent, and only the ``second" critical charge exists. Above the critical charge, the non-uniform strings have less entropy, as we see for $D\le 13$ near the GM point.

The sign change of $\sigma_2$ is an interesting phenomenon. 
In the present system, the charge works as a parameter that controls the order of transition between the uniform and non-uniform states. 
For $Q_{\mathrm{I,cr}}< Q < Q_{\mathrm{II ,cr}}$ in $D\le 13$, the stable phase of the non-uniform state emerges, realizing the second-order transition from the uniform phase.  
In addition to this, a second remarkable feature is that the non-uniform states become less stable near the GM point ($Q > Q_{\mathrm{II ,cr}}$). 
It is hard to provide a physical reason for these phenomena
\footnote{ 
The global thermodynamic consideration, in which the entropy is compared between a uniform black string and a localized black hole of same mass (and charge), provides us an intuitive guess for the critical dimension/charge (e.g. ~\cite{Sorkin:2004qq,Hovdebo:2006jy}). The present system, however, is out of the global argument since the corresponding localized black hole solution, if any, cannot couple to the gauge field of the same type.  
}. 
However, the second feature may be related to the fact that the non-uniform state itself vanishes at the GM point.

Next, let us focus on the Helmholtz free energy ($F=M-TS$) in a canonical ensemble, namely, the comparison of free energy between the non-uniform black string and a uniform black string of same temperature and charge. 
Such a difference is given by 
\begin{eqnarray}
 &&
	\frac{ F_{ \mathrm{NU} } - F_{ \mathrm{U} } }{ F_{ \mathrm{NU} } }
	 =
	\rho_2 \epsilon^4
	 +
	O( \epsilon^6 ),
 \\
 &&
 	\rho_2
	 =
 	- \frac{(d-3)(1-\Phi_{H0}^2)}{ 2[ 1 + (d-2)\Phi_{H0}^2] } 
		\biggl\{
  				\frac{   (d-1) \Phi_{H0}^2 \vartheta_1 }{ [ 1-(d-2)\Phi_{H0}^2] } 
    				\left[  \frac{ \vartheta_1 }{ d-3 } + 2  \tau_1
-  \frac{ 1 - (d-2)\Phi_{H0}^2 }{ (d-3)(1-\Phi_{H0}^2) } \Phi_{H1} 
   					 	\right]
 \nonumber
 \\
 &&
 \hspace{9cm}
  + \tau_1 \left[ s_1 + \frac{ (d-2) - \Phi_{H0}^2 }{ 1 - (d-2)\Phi_{H0}^2  } \tau_1  \right] \biggr\},
 \label{eq:rho2}
\end{eqnarray}
where we have used the first law of thermodynamics (Appendix \ref{appendix:entropy-diff}). Again the difference appears at $O(\epsilon^4)$.

The results are shown in Fig.~\ref{fig:6D-sigmarho} for $D=6,10,14$.
For the uncharged case,  $\rho_2$ is positive (for $D\le 12$), and thus the uniform state is more favorable due to the smaller free energy. 
As expected from the results of entropy comparison, the sign of $\rho_2$ changes as the charge increases. 
However, in this canonical ensemble, the sign changes only once.
$\rho_2$ tends to decease monotonically, and the critical charge is almost exactly the same as $Q_{\mathrm{I, cr}}$ for the entropy comparison. 
It is remarkable that the critical charge does not exist for $D\ge 13$ and $\rho_2$ is always negative.
This result can be understood with the general behavior of $\rho_2$ discussed above and with the fact that $\rho_2$ for $Q=0$ becomes negative for $D\ge 13$. 
Therefore, the non-uniform phase in the canonical ensemble for $D\ge 13$ is stable irrespective of the charge.

Finally, it is worthwhile to note that 
in Fig.~\ref{fig:6D-sigmarho} we have also shown the results obtained by taking $c_{0,1}(1)=+1, -1$, in addition to our standard choice $c_{0,1}(1)=0$. 
The three lines overlap completely, and we cannot distinguish between them. 
As we have discussed in Sec.~\ref{sec:zero-mode}, there is a subtlety in the choice of $c_{0,1}(1)$, but the present result means that our analysis is consistent and $c_{0,1}(1)$ does not affect the final results, as expected.

%%%%%%%%%%%%%%%%%%%%%%%%%%%%%%%%%%%%%%%%%%%%%%%%%%%%%%%%%%%%%%%%%%%%%%%%%%%%%%%
\subsection{Comparison with the criticality of uncharged system}
\label{sec:comp-d}
%%%%%%%%%%%%%%%%%%%%%%%%%%%%%%%%%%%%%%%%%%%%%%%%%%%%%%%%%%%%%%%%%%%%%%%%%%%%%%%

%%%%%%%%%%%%%%%%%%%%%%%%%%%%%%%%%%%%%%%%%%%%%%%%%%%%%%%%%%%%%%%%%%%%%%%%%%%%%%%
\begin{figure}[thbp]
	\begin{center}
	\includegraphics[width=8cm]{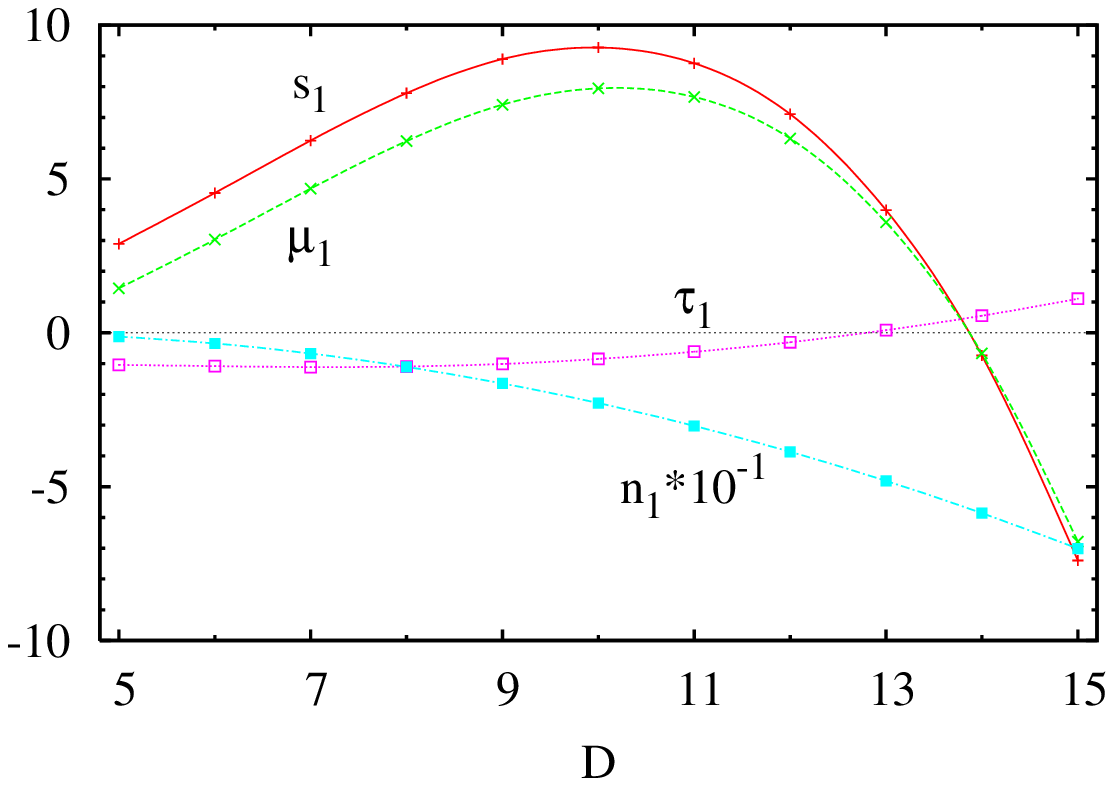}
	\includegraphics[width=8cm]{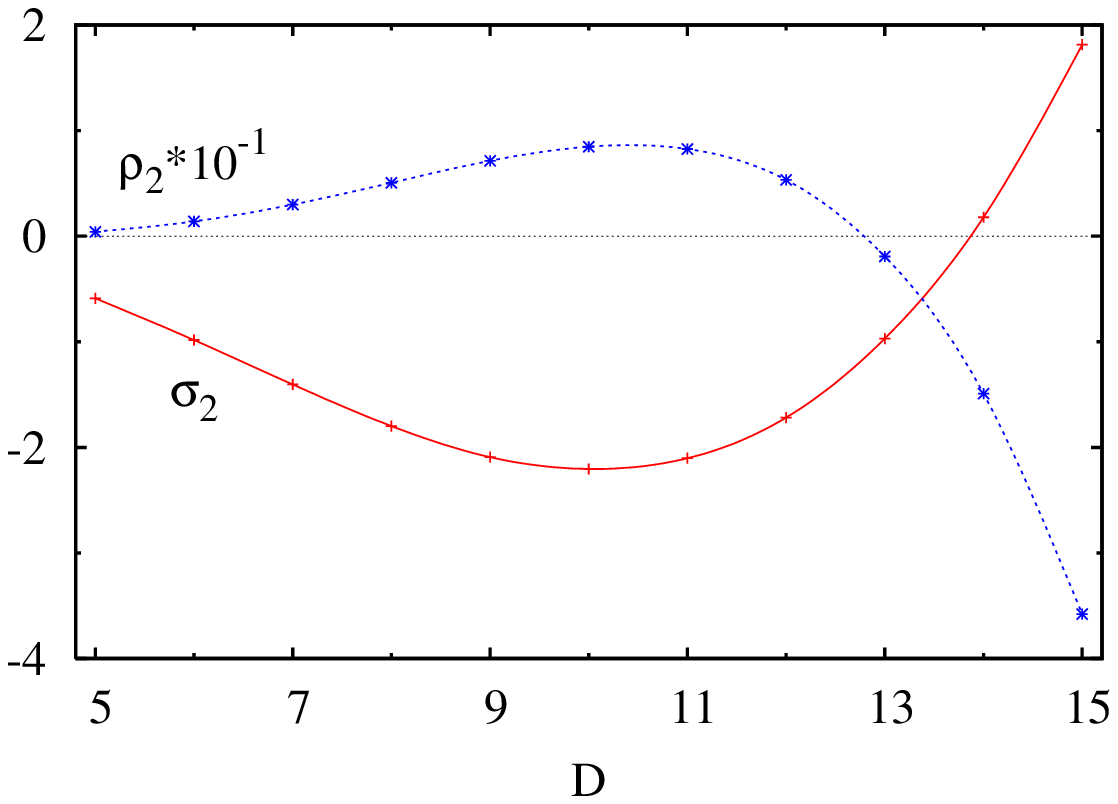}
\caption{
Dimensional dependence of physical quantities, e.g., $s_1, \mu_1$, etc., in addition to the entropy difference $\sigma_2$ and free energy difference $\rho_2$, for uncharged black strings ($Q=0$).
$s_1$ and $\mu_1$ become negative at $D=14$ and $\tau_1$ becomes positive at $D=13$. $n_1$ does not change its sign. 
$\sigma_2$ and $\rho_2$ change their sign at $D\approx 13.5$ and $D\approx 12.5$, respectively \cite{Sorkin:2004qq,Kudoh:2005hf}.
These results suggest that the phase transition from the uniform to non-uniform black strings in a microcanonical (canonical) ensemble is first order for $ 5 \leq D \leq 13 $ ($ 5 \leq D \leq 12$) and higher order for $ D \geq 14 $ ($ D \geq 13 $).
\label{fig:Neutral-sigmarho}
}
	\end{center}
\end{figure}
%%%%%%%%%%%%%%%%%%%%%%%%%%%%%%%%%%%%%%%%%%%%%%%%%%%%%%%%%%%%%%%%%%%%%%%%%%%%%%%

As we have already mentioned, we know that the stability of uncharged non-uniform string depends on the spacetime dimensions. 
(The number of extended transverse dimensions on tori $\mathrm{T}^p$ is not irrelevant \cite{Kol:2006vu},  and so we consider co-dimension one case of black string.) 
It is of great interest to compare the transition caused by the charge with the dimension-dependent transition of uncharged strings. 
In Fig.~\ref{fig:Neutral-sigmarho}, we show the dimensional dependence of physical quantities of uncharged non-uniform string. 
We also plot the entropy difference and free-energy difference in Fig.~\ref{fig:Neutral-sigmarho}.

Compared with Figs.~\ref{fig:6D-thermo s1 mu1 etc} and \ref{fig:6D-sigmarho}, we see large differences in their  qualitative behavior. The common behavior is that $s_1, \mu_1,\tau_1$ change their sign. For the charged case, the zero crossing of $s_1, \mu_1,\tau_1$ occurs almost simultaneously, in contrast to the uncharged case. 
Because of this feature, the critical charges in the canonical ensemble and the microcanonical ensemble are almost identical, while the critical dimensions for the two ensembles are different. 
A second striking disparity is that the charged system has a second critical charge.
This would be related to the fact that the GL static mode vanishes in the limit of
 $Q \to Q_{\mathrm{GM}}$ while the static mode survives in the limit of $D \gg 1$~\cite{Kol:2004pn}.

%%%%%%%%%%%%%%%%%%%%%%%%%%%%%%%%%%%%%%%%%%%%%%%%%%%%%%%%%%%%%%%%%%%%%%%%%%%%%%%

%%%%%%%%%%%%%%%%%%%%%%%%%%%%%%%%%%%%%%%%%%%%%%%%%%%%%%%%%%%%%
\section{Summary and discussion}
\label{sec:conclusion}
%%%%%%%%%%%%%%%%%%%%%%%%%%%%%%%%%%%%%%%%%%%%%%%%%%%%%%%%%%%%%

We have constructed the perturbatively non-uniform black strings (NUBS) coupled with the gauge field.
At the first order of perturbation, we have confirmed the realization of the correlated stability, i.e., the vanishing of the Gregory-Laflamme mode at the point where the background uniform black string (UBS) becomes thermodynamically stable.
We have pointed out that the emergence/vanishing of the static mode resembles phase transitions.
In fact, we have identified the critical exponent for the wavenumber of the static mode and found that the critical exponent is very close to $\beta=1/2$, which means a second-order phase transition. However, this result does not necessarily mean that the charged non-uniform black strings will arise from the uniform ones as a result of a second-order transition. 
This is because the black hole system is different in several aspects from the ordinary system in condensed matter \cite{Gubser:2001ac}, and, in addition, $k_0$ does not directly specify the state after the transition, but specifies only the point where such a state emanates.

The analysis of higher-order static perturbations allows us to study the charged non-uniform state with its (thermodynamic) stability. 
Our analysis shows that the charged non-uniform strings can have larger entropy than the corresponding uniform string in the appropriate range of charge (Fig. \ref{fig:6D-sigmarho}). This means that by adjusting the background charge the smooth transition (evolution) from the uniform state to non-uniform state becomes possible.  
This is in contrast to the uncharged system: we have known that uncharged black strings can have larger entropy (and smaller free energy) than that of uniform ones of the same mass (temperature) only in large spacetime dimensions, $D\geq 14$ ($D\geq 13$).

For the unchanged system, the phases of uniform and non-uniform states can be described by an analogous method of the Landau theory \cite{Kol:2006vu,Kol:2002xz}, where the external non-ordering field is the inverse of temperature. 
In the present case, the charge is a new non-ordering field, and because of this, phase diagrams should become multidimensional.
Such a system can be also described by the Landau theory. 
Consider the following Landau free energy:
\begin{eqnarray}
    f \sim  u_2 \phi^2 + u_4 \phi^4 + \cdots,
\end{eqnarray}
where the higher order terms maintaining stability are abbreviated. 
$u_2$ is a function of the non-ordering field, for instance, $u_2 \propto (T^{-1} - T^{-1}_{\mathrm{crit}})$ near the critical temperature in the uncharged case \cite{Kol:2006vu,Kol:2002xz}.
If $u_4$ is positive, a second-order transition is predicted, while there is a first-order transition if $u_4$ is negative. 
Like the normal-to-superfluid transition in liquid helium mixtures, $u_4$ can depend on non-ordering variables and its sign can vary. 
For the two non-ordering fields, the so-called tricritical point appears, at which three phases of disordered state and ordered states with first- and second-order transitions meet. 
Clearly, our charged system fits this argument. $u_4$ depends on charge (and temperature). 
It is an interesting issue to extend and refine the argument, based on \cite{Kol:2006vu,Kol:2002xz}. 
In the following, instead of pursuing its precise implementation, we discuss the $(M,n)$ phase diagram gathering all the information that we know.

\begin{figure}[tbp]
\begin{center}
\includegraphics[width=8cm]{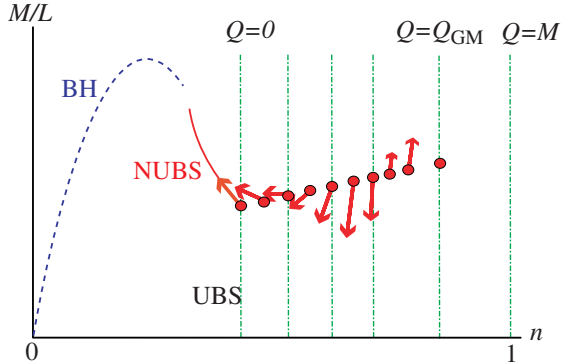}
\caption{
An illustration of charged non-uniform black strings (NUBS) in a $(M,n)$ plane, by projecting the charge axis on $(M,n)$ plane ($D\le 13$). 
The branch of uncharged NUBS and black hole are based on \cite{Kudoh:2004hs}.
Since the GL wavelength, and hence the bare mass, diverges in the limit $Q\to Q_{\mathrm{GM}}$, we plot the mass per unit length, $M/L$, where $L$ is the critical GL wavelength $L=2\pi/K$. 
The vertical dot-dashed lines represent uniform black strings with fixed changes, and the filled circles shows the sequence of critical GL points of a charged system.
Since there is no static mode for $Q>Q_{\mathrm{GM}}$, the sequence of the critical GL point terminates at $Q=Q_{\mathrm{GM}}$.
The lines with arrows show the direction in which the branch of charged NUBS emanates. 
The direction of arrows can be read off from Fig.~\ref{fig:6D-thermo s1 mu1 etc}.
\label{fg:Mn-diagram} 
}
	\end{center}
\end{figure}

A mass-tension $(M,n)$ diagram is commonly used to classify the black objects in the KK compactification. 
We project out the degree of freedom of charge on the $(M,n)$-plane and consider the phase on this plane. 
In Fig.~\ref{fg:Mn-diagram}, we illustrate a $(M,n)$ diagram for a charged system ($D\le 13$). 
Although $L \propto 1/k_0$ diverges in the GM limit ($Q=Q_{\mathrm{GM}}$), the mass per unit length remains finite, so that we adopt it in the figure. 
The sequence of the marginally stable charged uniform black strings (UBS) is given by Eqs. (\ref{eq:mass}) and (\ref{eq:tension}), and the non-uniform branch can be read off from Fig.~\ref{fig:6D-thermo s1 mu1 etc}.
Corresponding to $\mu_1<0$ in $Q_{\mathrm{I,cr}}<Q <Q_{\mathrm{II,cr}}$, the branch of NUBS declines in the region. However, the corrections to the mass and tension become $\mu_1>0$, $n_1>0$, and then the branch of NUBS ascends near the GM point. 
These behaviors show that the phase space of black hole and string has a very rich structure.

Basing our ideas on this phase structure, let us speculate about a possible dynamical evolution of sufficiently charged black strings with $\sigma_2>0$.
First of all, we adopt the plausible assumption that the conservation of magnetic charge prevents pinching-off uniform black strings/branes.
(This is also expected from the fact that the black hole cannot couple to the same type of form field.)
Thus a possible scenario is that once the uniform black string reaches the GL point, via radiation, etc., the unstable critical string will transit to a non-uniform black string adiabatically if the charge is appropriately chosen so that the non-uniform branch is less massive and entropically favored.
At present, the dynamical stability of non-uniform black stings is not known. 
However, it is possible that a shrinking part of the wavy (apparent) horizon will locally evolve into the extremality, and the stabilization works locally, making the whole black string stabilized in the end.
Therefore, a wavy black string would be a possible end state in such a dynamical evolution. 
To identify the sequence of evolution, it will be useful to construct charged non-uniform black strings non-perturbatively. 
Since there is no corresponding branch of black holes carrying the same type of charge, we have no reason to expect that the branch of black string remains near the uncharged one even for a weak charge.
Therefore, it is interesting to know how the branch deviates from that of an uncharged string at large non-uniformity.

So far, we have discussed the phase space in $Q< Q_{\mathrm{GM}}$ where the static modes exist.
However, the phase space in $Q > Q_{\mathrm{GM}}$ may be also interesting to explore. 
An argument proposed by Horowitz and Maeda~\cite{Horowitz:2002ym} is that (stable) non-uniform  black strings would exist near the extremality. 
Such a solution has not been discovered so far, and it will be interesting to know how the branches of near-extremal non-uniform black strings emerge and how they are embedded in the diagram. 
However, we should recall here that the scenario of such new branch is possible only when cycles on the horizon cannot shrink to zero size.
So if it can pinch off, a near-extremal non-uniform black string is not an inevitable result.

Finally, let us comment on possible extensions of the work.
For simplification, we have not taken into account the dilaton, but it will be straightforward to include the dilaton and other types of gauge fields.
As discussed above, it is important to figure out the phase structure in fully non-linear regimes in the present system.
This problem will be addressed in our forthcoming paper. 
The dynamical evolution of GL instability in a charged system will be the most challenging problem.
To see what happens dynamically when the charge crosses the critical value will be quite fascinating. 
The dynamics of a string's evolution may change drastically at the critical charge. 
This can be achieved by a straightforward extension of the analysis in \cite{Choptuik:2003qd}, and the result will also provide an expectation for the dynamical evolution of uncharged black strings above the critical dimension ($D \ge 14$).

%%%%%%%%%%%%%%%%%%%%%%%%%%%%%%%%%%%%%%%%%%%%%%%%%%%%%%%%%%%%%%%%%%%%%%%%%%%%%%%
%%%%%%%%%%%%%%%%%%%%%%%%%%%%%%%%%%%%%%%%%%%%%%%%%%%%%%%%%%%%%%%%%%%%%%%%%%%%%%%

%%%%%%%%%%%%%%%%%%%%%%%%%%%%%%%%%%%%%%%%%%%%%%%%%%%%%%%%%%%%%%%%%%%%%%%%%%%%%%%
\section*{Acknowledgements} 
%%%%%%%%%%%%%%%%%%%%%%%%%%%%%%%%%%%%%%%%%%%%%%%%%%%%%%%%%%%%%%%%%%%%%%%%%%%%%%%
We would like to thank to K.~-I.~Maeda, B.~Kol,  E.~Sorkin, T.~Wiseman, S. Mukohyama, G.~Horowitz and S. Tomizawa for useful discussion and comments. 
This work is supported in part by a Grant for The 21st Century COE Program (Holistic Research and Education Center for Physics Self-Organization Systems) and a Grant for Special Research Projects at Waseda University.
HK is supported by the JSPS. 
HK is grateful to the KITP and the workshop ``Scanning New Horizons: GR Beyond 4 Dimensions'' in Santa Barbara for the stimulating environment. 
%%%%%%%%%%%%%%%%%%%%%%%%%%%%%%%%%%%%%%%%%%%%%%%%%%%%%%%%%%%%%%%%%%%%%%%%%%%%%%%

%%%%%%%%%%%%%%%%%%%%%%%%%%%%%%%%%%%%%%%%%%%%%%%%%%%%%%%%%%%%%%%%%%%%%%%%%%%%%%%
\appendix
%% \setcounter{equation}{0}
%%%%%%%%%%%%%%%%%%%%%%%%%%%%%%%%%%%%%%%%%%%%%%%%%%%%%%%%%%%%%%%%%%%%%%%%%%%%%%%

%%%%%%%%%%%%%%%%%%%%%%%%%%%%%%%%%%%%%%%%%%%%%%%%%%%%%%%%%%%%%%%%%%%%%%%%%%%%%%%
\section{Einstein equations in static axisymmetric spacetimes}
\label{appendix:Ricci}
%%%%%%%%%%%%%%%%%%%%%%%%%%%%%%%%%%%%%%%%%%%%%%%%%%%%%%%%%%%%%%%%%%%%%%%%%%%%%%%

We present the Ricci tensor components for the general metric ansatz (\ref{eq:conformal}). Here, we denote the partial derivatives by $\partial_r A = A'$ and $\partial_z A = \dot{A}$. In addition, we specify the coordinates of the angular part by
$d\Omega_{d-2}^{2}
=
d\theta_1^2
+
\sin^2\theta_1^2 d\theta_2^2
+
\cdots
+
\sin^{2}\theta_1
\cdots
\sin^{2}\theta_{d-3} d\theta_{d-2}^{2}.
$
The Ricci tensor is calculated as follows:
%%%%%%%%%%%%%%%%%%%%%%%%%%%%%%%%%%%%%%%%%%%%%%%%%%%%%%%%%%%%%
\begin{eqnarray}
&&
R_{tt}
 =
e^{2A - 2B - 2H}
	\biggl\{
			e^{2 H}
   			\dot{A} 
   				\left[ \dot{A} + (d-2) \dot{C} \right]
   				 +
   			 A' \left[ A' - H'  + (d-2) \left( C' + \frac{1}{r} \right) \right] 
   			 +
   			e^{2 H} \ddot{A} + A'' 
	\biggr\},
\cr
%%%%%%%%%%%%%%%%%%%%%%%%%%%%%%%%%%%%%%%%%%%%%%%%%%%%%%%%%%%%%
&&R_{rr}
=
\frac{(d-2) H' }{r}-e^{2 H}  \dot{B}    
   \left[ \dot{A} + (d-2) \dot{C} \right]
 + A' \left(H' -A' +B' \right) 
\cr
&&
\hspace{1cm}
 + (d-2) C' \left(H' - C' -\frac{2}{r} \right)
    + B' \left[ H' +(d-2) \left( C' + \frac{1}{r} \right) \right] 
   - [ A  + B + (d-2) C ]'' 
   - e^{2 H} \ddot{B}  ,
\cr
&&R_{zz}
=
\dot{A} 
   ( \dot{B } - \dot{A}) + (d-2) \dot{C} (\dot{B} -\dot{C})
\nonumber
\\
&&
\hspace{2cm}  
   + e^{-2 H} B'
   \left[ H' -A' -(d-2) \left(C' + \frac{1}{r} \right) \right]
   - \ddot{A} - \ddot{B} -(d-2) \ddot{C}
   - e^{-2 H} B''  ,
\cr
%%%%%%%%%%%%%%%%%%%%%%%%%%%%%%%%%%%%%%%%%%%%%%%%%%%%%%%%%%%%%
&&R_{\theta_1\theta_1}
=
r^2 e^{-2 B +2 C -2 H}  
\biggl\{
    \frac{H'}{r} - \frac{ A'}{r} 
   - \frac{(d-3) \left[1-e^{2(B - C +H)}\right]}{r^2}
   -e^{2 H} \dot{C} 
   \left[ \dot{A} +(d-2) \dot{C} \right] 
\cr
&& 
\hspace{2cm}
   + C' \left[\frac{4-2d}{r} + H' 
   - A' -(d-2) C' \right]
     - e^{2 H}\ddot{C} - C''
\biggr\},
\cr
%%%%%%%%%%%%%%%%%%%%%%%%%%%%%%%%%%%%%%%%%%%%%%%%%%%%%%%%%%%%%
&&R_{rz}
=
 ( \dot{B} - \dot{A} )
 A' + \left[ \dot{A} + (d-2) \dot{C} \right] B' 
 +(d-2) \left( \dot{B}  -\dot{C} \right)
 \left( C' + \frac{1}{r} \right)
 - \dot{A}' -(d-2) \dot{C}',
%%%%%%%%%%%%%%%%%%%%%%%%%%%%%%%%%%%%%%%%%%%%%%%%%%%%%%%%%%%%%
\label{eq:ricci}
\end{eqnarray}
and the other angular components are given by
$
 R_{\theta_1}^{\theta_1}
=
 R_{\theta_2}^{\theta_2}
=
\cdots
=
 R_{\theta_{d-2}}^{\theta_{d-2}}.
$
%%%%%%%%%%%%%%%%%%%%%%%%%%%%%%%%%%%%%%%%%%%%%%%%%%%%%%%%%%%%%
For the magnetic field given by $F=Q \varepsilon_{d-2}$, the non-zero components of the RHS of the Einstein equation (\ref{eq:form-EOM}), which we denote by $S_{\mu\nu}$, are
\begin{eqnarray}
&&
	S_t^t
	 =
	S_r^r
	 =
	S_z^z
	 =
	-\frac{ d-3 }{ 2(d-1) } \frac{ Q^2 }{ r^{ 2(d-2) } e^{ 2(d-2) C } },
 \cr
 &&
 	S_{ \theta_{i} }^{ \theta_{i} }
	 =
	\frac{1}{d-1} \frac{ Q^2 }{ r^{ 2(d-2) } e^{ 2(d-2)C } }
	 ,\;\;
	( i=1,2, \ldots, d-2 ). 
 \label{eq:Einstein-source}
\end{eqnarray}

%%%%%%%%%%%%%%%%%%%%%%%%%%%%%%%%%%%%%%%%%%%%%%%%%%%%%%%%%%%%%%%%%%%%%%%%%%%%%%%
\section{Other form field components}
\label{appendix:electric}
%%%%%%%%%%%%%%%%%%%%%%%%%%%%%%%%%%%%%%%%%%%%%%%%%%%%%%%%%%%%%%%%%%%%%%%%%%%%%%%
In this paper, we focus only on the angular parts of the form field $F$.
In $d=4$ and $d=5$, however, the form field can have other components, corresponding to an electric field.
In this section, we briefly discuss these components compatible to the conformal ansatz (\ref{eq:conformal}).

For $d=5$, $F_{trz}$ component can exist in addition to (\ref{eq:general-F}).
We put this part as
\begin{eqnarray}
	F_{e}
	 =
	E(r,z) \;\omega_{t} \wedge \omega_{r} \wedge \omega_{z},
\end{eqnarray}
where
$\omega_{t} = e^{A} dt $,
$ \omega_{r} = e^{B+H} dr $
and
$ \omega_{z} = e^{B} dz $.
Substituting this into EOM (\ref{eq:form-EOM}), we obtain a solution
\begin{eqnarray}
	E(r,z) = \frac{ Q_{e} }{ r^3 e^{3C} }, 
 \label{eq:6D-electric}
\end{eqnarray}
where $ Q_{e} $ is an integration constant.
The solution (\ref{eq:6D-electric}) is the general solution of electric field, which we ignored in our analysis.
Note that since the 3-form is self-dual in 6 dimensions, the electric and magnetic parts contribute to the Einstein equation in a similar manner: if we consider the electric and magnetic part simultaneously, the source term of the Einstein equation is obtained just by replacing the magnetic charge $Q^2$ in (\ref{eq:Einstein-source}) by $ Q^2 + Q_e^2 $.
In other words, the electric field in $d=5$ can be analyzed in parallel with the magnetic one, and therefore the self-duality does not affect the final results.

%It is noted, however, that since the 3-form is self-dual in 6 dimensions, the electric and magnetic parts contribute to the Einstein equation in a similar manner: if we consider the electric and magnetic part simultaneously, the source term of the Einstein equation is obtained just by replacing the magnetic charge $Q^2$ in (\ref{eq:Einstein-source}) for $ Q^2 + Q_e^2 $.

For $ d=4 $, the electric components in addition to (\ref{eq:general-F}) are possible and they are given by
\begin{eqnarray}
	F_{e}
	 =
	E_{r}(r,z) \; \omega_{t} \wedge \omega_{r}
	 +
	E_{z}(r,z) \; \omega_{t} \wedge \omega_{z}
	 +
	E_{rz}(r,z) \; \omega_{r} \wedge \omega_{z}.
\end{eqnarray}
From the EOMs, we find 
$
	E_{r z} = Q_{rz}/ (r^2 e^{2C+A} ),
$ where $Q_{rz}$ is an integration constant. 
This is the general solution of $r$-$z$ component.
The other components are given by solving the EOM and Bianchi identities:
\begin{eqnarray}
	( \partial_{r} + 2 r^{-1} ) ( e^{ B+2C }  E_r )
	 +
	e^H \partial_{z} ( e^{ B+2C } E_z )
	 &=&
	0,
 \nonumber
 \\
	\partial_{r} ( e^{ A+B } E_z )
	 -
	e^{H} \partial_{z} ( e^{ A+B } E_r )
	 &=&
	0.
 \label{eq:PDEs}
\end{eqnarray}

%%%%%%%%%%%%%%%%%%%%%%%%%%%%%%%%%%%%%%%%%%%%%%%%%%%%%%%%%%%%%%%%%%%%%%%%%%%%%%%
\section{Specific heat and chemical potential}
\label{appendix:capacity}
%%%%%%%%%%%%%%%%%%%%%%%%%%%%%%%%%%%%%%%%%%%%%%%%%%%%%%%%%%%%%%%%%%%%%%%%%%%%%%%

We derive the formula for the specific heat used in Sec. \ref{sec:Physical quantities}.
First of all, note that every thermodynamical quantity of uniform black strings ($a = b = c = 0$) can be written as functions of $r_{+}$ and $q$.
From Eq.~(\ref{eq:charge}), we have
\begin{eqnarray} 
	q 
	 =
	\frac{1}{r_+^2}
		\left[
			\frac{ 16 \pi G_{ d+1 } }{ (d-1) L \Omega_{d-2} } ~ Q
		\right]^{ 2/(d-3) }. 
 \label{eq:q by Q and r_+}
\end{eqnarray}
Substituting (\ref{eq:q by Q and r_+}) into the mass formula (\ref{eq:mass}) or the product of $S$ and $T$, one finds
\begin{eqnarray}
	r_+ ^{d-3} 
	 &=&
	\frac{ 16 \pi G_{d+1} }{ L \Omega_{d+1} }  \frac{ M }{ 2(d-2) } 
 		\left(
 			1 + \sqrt{ 1 - \frac{ 4(d-2) }{ (d-1)^2 } \frac{ Q^2 }{ M^2 } }
 		\right)
 	\nonumber
 	\\
	 &=&
	\frac{ 16 \pi G_{d+1} }{ L \Omega_{d+1} } 
	\frac{ ST }{ 2(d-3) }
		\left(
			1 + \sqrt{ 1 + \frac{ 4(d-3) }{ d-1 } \frac{ Q^2 }{ ST } }
		\right) . 
 \label{eq:r_+ by MTS}
\end{eqnarray}

Let us first discuss the specific heat. 
Utilizing (\ref{eq:q by Q and r_+}) and (\ref{eq:r_+ by MTS}), the RHS of the temperature (\ref{eq:temperature}) can be written in terms of $ Q$ and ${S T}$, i.e., $ T= T(  Q,  S  T)$. 
Differentiating this equation with respect to $ S$, we easily find $ (\partial{  S}/ \partial{ T} )_Q$, and the specific heat is evaluated by
\begin{eqnarray}
	 C_{Q} 
	  \equiv 
	 \left( \frac{ \partial M }{ \partial T } \right)_{Q} 
	  =
	 T \left( \frac{ \partial S }{ \partial T } \right)_{Q}, 
 \label{eq:derivation of capacity}
\end{eqnarray}
which yields the result (\ref{eq:def capacity}).

The method explained above is the most straightforward for obtaining thermodynamic quantities, e.g., the specific heat. 
However, it is sometimes difficult to get an explicit expression of a thermodynamic quantity in terms of relevant thermodynamic variables. 
In such cases, it is necessary to perform rather messy calculations. 
Instead of such a direct method, we explain a systematic method to calculate thermodynamic variables which have parametric representations.

As an illustration, we begin with the calculation of chemical potential.
The chemical potential is given by  
\begin{eqnarray}
	\Phi_H
	 \equiv
	\left( \frac{ \partial M }{ \partial Q } \right)_{S}
	 =
	\left( \frac{ \partial M }{ \partial r_{+} } \right)_q
	\left( \frac{ \partial r_{+} }{ \partial Q } \right)_S
	 +
	\left( \frac{ \partial M }{ \partial q } \right)_{ r_+ }
	\left( \frac{ \partial q }{ \partial Q } \right)_S.
 \label{eq:potential-calc}
\end{eqnarray}
To calculate $(\partial r_{+}/\partial Q)_{S}$ and $(\partial q/\partial Q)_{S}$, the charge $Q(r_+,q)$ is brought into the form of $ Q = $ $ Q( r_{+}, q(S, r_{+}) ) $ $ = Q(r_{+}(S, q), q) $.
Differentiating these expressions with respect to $r_{+}$ and $q$ as 
\begin{eqnarray}
 &&
	\left(	\frac{ \partial Q }{ \partial r_+ } \right)_{S}
	 =
	\left( \frac{ \partial Q }{ \partial r_+ } \right)_{q}
	 +
	\left( \frac{ \partial Q }{ \partial q } \right)_{r_+}
	\left( \frac{ \partial q }{ \partial r_+ }\right)_{S} ,
 \nonumber
 \\
 &&
 	\left( \frac{ \partial Q }{ \partial q } \right)_{S}
	 =
	\left( \frac{ \partial Q }{ \partial q } \right)_{r_+}
	 +
	\left( \frac{ \partial Q }{ \partial r_+ } \right)_{q}
	\left( \frac{ \partial r_+ }{ \partial q } \right)_{S},
 \label{eq:dQdr}
\end{eqnarray}
and substituting them into (\ref{eq:potential-calc}), we finally obtain $\Phi_{H}=q^{(d-3)/2}$.
By replacing variables suitably in the above calculation of chemical potential, we can easily calculate the specific heat $C_Q$.
%and isothermal permittivity $\epsilon_T$.

%%%%%%%%%%%%%%%%%%%%%%%%%%%%%%%%%%%%%%%%%%%%%%%%%%%%%%%%%%%%%%%%%%%%%%%%%%%%%%%
\section{Entropy and free energy difference}
\label{appendix:entropy-diff}
%%%%%%%%%%%%%%%%%%%%%%%%%%%%%%%%%%%%%%%%%%%%%%%%%%%%%%%%%%%%%%%%%%%%%%%%%%%%%%%

We use the expansion (\ref{eq:expansion of MSTQF}) again for all relevant physical quantities. The entropy difference between the non-uniform and uniform black strings for same mass and charge is given by
\begin{eqnarray}
	\frac{ S_{ \mathrm{NU} } - S_{ \mathrm{U} } }
		 { S_{ \mathrm{U} } }
 && 
	 = 
	\frac{ \sum_{p=0}^{\infty} S_{p} \epsilon^{2p} }{ S_{0} + \Delta S } - 1
 \cr
 &&
	 = 
	\epsilon^2 
	\left[  \frac{ S_{1} }{ S_{0} } -\frac{ \Delta S^{(1)} }{ S_0 } \right] 
	 +
	\left[
		\frac{ S_2 }{ S_0 } 
		 -
		\frac{ \Delta S^{(2)} }{ S_0 }  
		 -
		\frac{ S_1 }{ S_0 } \frac{ \Delta S^{(1)} }{ S_{0} } 
		 +
		\left( \frac{ \Delta S^{(1)} }{ S_0 } \right)^2 
	\right]
		\epsilon^4
		+
	O( \epsilon^6 ),
 \label{eq:entropy-diff}
\end{eqnarray}
where $\Delta S = \sum_{p=1} \Delta S^{(p)} \epsilon^{2p}$ is the difference of entropy between the critical uniform black string and the \textit{uniform} black string that has the same mass and charge as the non-uniform black string.

$S_{1}/S_{0}$ and $S_{2}/S_{0}$ in (\ref{eq:entropy-diff}) are easily computed from the first law for a fixed asymptotic length of the circle,
$\mathrm{d} M = T \mathrm{d} S  +  \Phi_H \mathrm{d} Q$. 
Integrating the first law with the expansion, we obtain
\begin{eqnarray}
 &&
 	M_{1}
 	 =
 	T_0 S_1 + \Phi_{H 0} Q_{1},
 \nonumber
 \\
 &&
 	M_{2}
 	 =
 	T_0 S_2
 	 +
 	\frac{1}{2} T_1 S_1
 	 +
 	\Phi_{H 0} Q_{2}
 	 +
 	\frac{1}{2} \Phi_{H1} Q_1.
 \label{eq:mass-diff}
\end{eqnarray}
Furthermore, from formulae (\ref{eq:mass}), (\ref{eq:temperature}), (\ref{eq:entropy}), and (\ref{eq:charge}), we can find that the mass of uniform black string $M_0$ can be written in two ways:
\begin{eqnarray}
	M_{0}
	 =
	\frac{ d-2 + q^{d-3} }{ (d-3)(1-q^{d-3})} T_0 S_0
	 =
	\frac{ d-2 + q^{d-3} }{ (d-1)q^{ (d-3)/2 } } Q_0.
 \label{eq:mass-diff2}
\end{eqnarray}
The first equality can be also confirmed by the Smarr formula.
From (\ref{eq:mass-diff}) and (\ref{eq:mass-diff2}), we can write
$S_1/S_0$ and $S_2/S_0$
in terms of
$M_1/M_0$,
$M_2/M_0$,
$Q_1/Q_0$,
$Q_2/Q_0$,
$\Phi_{H1}/\Phi_{H0}$
and
$T_1/T_0$.

Here, we focus on the quantity $\Delta S/S_0$ in (\ref{eq:entropy-diff}). 
As mentioned before, $\Delta S$ is the entropy change of the uniform black string due to the change of mass and charge.
Therefore, $\Delta S$ can be computed by
\begin{eqnarray}
	&&
	\Delta S 
	 \simeq
	\left( \frac{ \partial S }{ \partial M } \right)_{Q } \Delta M 
	 +
	\left( \frac{ \partial S }{ \partial Q } \right)_{M } \Delta Q
	\nonumber
	\\
	&&
	\hspace{3cm}
	 +
	\frac{1}{2}
		\left( \frac{ \partial^2 S }{ \partial M ^2 } \right)_{Q } \Delta M^2
	 +
	\frac{1}{2}
		\left( \frac{ \partial^2 S }{ \partial Q ^2 } \right)_{M } \Delta Q^2
	 + 
 	\left( \frac{ \partial^2 S }{ \partial Q  \partial M} \right) \Delta Q \Delta M,
 \label{eq:Delta S-expansion}
\end{eqnarray}
where $S$, $M$, and $Q$ are for uniform strings $(a=b=c=0)$.
One can obtain the explicit dependence of $S$ on $M$ and $Q$ by substituting the expressions of $q$ and $r_+$, Eqs. (\ref{eq:q by Q and r_+}) and (\ref{eq:r_+ by MTS}), into Eq.~(\ref{eq:mass}). But, it seems to be easiest to calculate (\ref{eq:Delta S-expansion}) as follows. 
The partial derivatives of $S$ with respect to $M$ and $Q$ in Eq.~(\ref{eq:Delta S-expansion}) are given by 
\begin{eqnarray}
	\left( \frac{ \partial  }{ \partial M } \right)_{Q}
	 &&
	 =
	\left( \frac{ \partial r_+ }{ \partial M } \right)_{Q}
	\left( \frac{ \partial     }{ \partial r_+ } \right)_q
	 +
	\left( \frac{ \partial q }{ \partial M } \right)_{Q}
	\left( \frac{ \partial   }{ \partial q } \right)_{r_+} , 
 \nonumber
 \\
	\left( \frac{ \partial }{ \partial Q } \right)_{M}
	 &&
 	 =
	\left( \frac{ \partial r_+ }{ \partial Q } \right)_{M}
	\left( \frac{ \partial     }{ \partial r_+} \right)_q
	 +
	\left( \frac{ \partial q }{ \partial Q } \right)_{M}
	\left( \frac{ \partial   }{ \partial q } \right)_{r_+} . 
 \label{eq:pd}
\end{eqnarray}
From Eq. (\ref{eq:charge}), $r_+$ and $q$ are rewritten as
$r_{+}=r_{+}(Q, q)$
and
$q=q(Q, r_{+})$,
so
$M(r_+, q)$
can be written as
$M=M( r_{+}(Q, q), q ) = M( r_{+}, q(Q, r_+) )$. 
Differentiating these relations with respect to $r_{+}$ and $q$, we obtain the coefficient in Eq.~(\ref{eq:pd}).
Then, by setting
$\Delta M = M_1 \epsilon^2 + M_2 \epsilon^4$
and
$\Delta Q = Q_1 \epsilon^2 + Q_2 \epsilon^4$,
we obtain
$\Delta S = \Delta S^{(1)} \epsilon^2 + \Delta S^{(2)} \epsilon^4$.

We can find the leading term to be the same as the one for the non-uniform black string:
\begin{eqnarray}
	\frac{ \Delta S^{(1)} }{ S_0 } 
 	 =
 	\delta S_{1}
  	 =
  	\frac{ d-2 + {\Phi_{H 0}^2} }{ (d-3) (1- {\Phi_{H 0}^2} ) }  \delta M_1
	 -
	\frac{ (d-1){\Phi_{H 0}^2} }{ (d-3) (1-{\Phi_{H 0}^2} ) }  \delta Q_1,
 \label{eq:s_1/s_0 by M1Q1}
\end{eqnarray}
where we again abbreviate
$\delta Y_1 = Y_{1}/Y_{0}$
($Y = S, \; M, \;Q$).
Hence the term of order $O(\epsilon^2)$ vanishes in the entropy difference and the leading order correction comes from $O(\epsilon^4)$.  
Note that $M_2/M_0$ and $Q_2/Q_0$ cancel out in the final expression as expected from the cancellation at the order $O(\epsilon^2)$. 
In the end the entropy difference for the same mass and charge is given by 
\begin{eqnarray}
%%
%%=======  Result written by M_1 and Q_1 (Don't delete) ==============
%% \frac{S_{\mathrm{NU}}-S_{\mathrm{U}}}{S_{\mathrm{U}}}
%% =
%%- \frac{ \left( d-2 + {\mathcal Q} \right)}
%%{2 (d-3)(1-{\mathcal Q} ) }  \delta M_1
%% \left[ \frac{\left(  d-2 + {\mathcal Q}\right) 
%%             ( 1 - ( d-2) {\mathcal Q})   }{(d -3)  
%%             ( 1 -{\mathcal Q} )   \left( d-2 - {\mathcal Q} \right) }\delta M_1
%%        + \delta T_1    \right]  
%% \cr
%%&&  +    \frac{ (d -1) {\mathcal Q} }{2 (d-3) ( 1 - {\mathcal Q}) }
%%     \delta Q_1
%%     \biggl\{
%%       \delta T_1 
%%       + \delta Q_1 \frac{ [ (d -3)(d -2) + (d -5){\mathcal Q}
%%              + 2 {\mathcal Q}^2 ]}{(d-3) (1-{\mathcal Q}) (d-2 -{\mathcal Q})}  
%%\cr
%% &&  -2   \delta M_1  \frac{ [(d-4)(d-2) + 2(d-3){\mathcal Q} 
%%             + {\mathcal Q}^2  ] 
%%         }{(d-3) (1-{\mathcal Q}) (d-2 - {\mathcal Q})}  
%%  \biggr\}, 
%%
%%  ${\mathcal Q} = q^{d-3}$
%%=============================================================
%%
%%=======  Result written by S_1 and Q_1 (Don't delete) ==============
%%
 &&
 	\frac{ S_{ \mathrm{NU} } - S_{\mathrm{U}} }{ S_{ \mathrm{U} } }
 = 
 	\frac{ (d-1){\Phi_{H 0}^2} ~ \delta Q_1 }{ d-2- {\Phi_{H 0}^2} }  
 		\left\{ 
   			\frac{ d-2 }{ 2(d-3) } 
   			\left[
   				\delta Q_1 - \frac{ d-2-{\Phi_{H 0}^2} }{ (d-2)(1- {\Phi_{H 0}^2}) }
   				\delta \Phi_{H1}
   			\right]
   			-  \delta S_1
 		\right\} \epsilon^4
 	 \nonumber
 	 \\
 	 &&
 	 \hspace{7cm}
  	 -
  	\frac{\delta S_1}{2}
    	\left[
  			\frac{ 1 - (d-2) {\Phi_{H 0}^2} }{ d-2 - {\Phi_{H 0}^2} } \delta S_1
  			 +
  			\delta T_1
		\right]\epsilon^4 , 
\end{eqnarray}
where we have used the relation (\ref{eq:s_1/s_0 by M1Q1}) to simplify the result.
Last, by replacing $\delta Y_{1}$ for invariant quantities, we have Eq.~(\ref{eq:sigma2}).

The derivation of the formula for the free-energy comparison is possible in a similar way to that of entropy. First of all, the free energy ($F=M-TS$) of a {\it uniform} solution can be written in two ways from Eqs. (\ref{eq:mass}), (\ref{eq:temperature}), (\ref{eq:entropy}), and (\ref{eq:charge}):
\begin{eqnarray}
	F_{0}
	 =
	\frac{ 1 + (d-2)q^{d-3} }{ (d-3)(1-q^{d-3}) } T_0 S_0 
	 =
	\frac{ 1 + (d-2)q^{d-3} }{ (d-1)q^{(d-3)/2} } Q_0.
 \label{eq:F_0 by TSQ_0}
\end{eqnarray}
The difference of free energy between uniform and non-uniform black strings is given by
\begin{eqnarray}
 && 
	\frac{ F_{ \mathrm{NU} } - F_{ \mathrm{U} } }{ F_{ \mathrm{U} } }
 	 =
	\frac{ \sum_{p=0}^{\infty} F_{p} \epsilon^{2p} }{ F_{0} + \Delta F } - 1,
 \label{eq:free-energy-diff}
\end{eqnarray}
where $\Delta F $ is the free-energy change of a \textit{uniform} black string due to the change of temperature and charge, which we denote by $\Delta T $ and $\Delta Q $, respectively.
From the definition of the free energy and the first law, we obtain
$ \mathrm{d} F	= -S \mathrm{d} T + \Phi_H  \mathrm{d} Q $. 
Integrating this by using the expansion (\ref{eq:expansion of MSTQF}), 
\begin{eqnarray}
 &&
 	F_1
 	 =
 	-S_0 T_1 + {\Phi_{H 0} } Q_1,
 \nonumber
 \\
 &&
 	F_2
 	 =
 	- S_0 T_2
 	- \frac{1}{2} S_1 T_1
 	+ {\Phi_{H 0} } Q_2 + \frac{1}{2} \Phi_{H1} Q_{1}.
 \label{eq:F_1,F_2}
\end{eqnarray}
Using (\ref{eq:F_0 by TSQ_0}) and (\ref{eq:F_1,F_2}), $F_1/F_0$ and $F_2/F_0$ are written in terms of $T_1/T_0$, $T_2/T_0$, $Q_1/Q_0$, $Q_2/Q_0$, $\Phi_{H1}/\Phi_{H0}$, and $S_1/S_0$.

Now, we focus on the term of $\Delta F/F_0$ in (\ref{eq:free-energy-diff}). 
A necessary correction of the free energy to the uniform string is 
\begin{eqnarray}
 &&	
	\Delta F 
	 \simeq
	\left( \frac{ \partial F }{ \partial T } \right)_{Q} \Delta T 
	 +
	\left( \frac{ \partial F }{ \partial Q } \right)_{T} \Delta Q
 \nonumber
 \\
 &&
 \hspace{3cm}
	+ \frac{1}{2}
		\left( \frac{ \partial^2 F }{ \partial T ^2 } \right)_{Q } \Delta T^2
	+ \frac{1}{2}
		\left( \frac{ \partial^2 F }{ \partial Q ^2} \right)_{T } \Delta Q^2
	+   \left( \frac{ \partial^2 F }{ \partial Q  \partial T } \right) \Delta Q \Delta T,
 \label{eq:Delta F-expansion}
\end{eqnarray}
where the differential operators acting on $F$ are evaluated by using the following relation:
\begin{eqnarray}
	\left( \frac{ \partial }{ \partial T } \right)_{Q}
	 &=&
	\left( \frac{ \partial r_+ }{ \partial T   } \right)_{Q}
	\left( \frac{ \partial     }{ \partial r_+ } \right)_q
	 +
	\left( \frac{ \partial q   }{ \partial T   } \right)_{Q}
	\left( \frac{ \partial     }{ \partial q   } \right)_{r_+} , 
	 \cr
	\left( \frac{ \partial     }{ \partial Q   } \right)_{T}
	 &=&
	\left( \frac{ \partial r_+ }{ \partial Q   } \right)_{T}
	\left( \frac{ \partial     }{ \partial r_+ } \right)_q
	 +
	\left( \frac{ \partial q   }{ \partial Q   } \right)_{T}
	\left( \frac{ \partial     }{ \partial q   } \right)_{r_+} . 
\end{eqnarray}
Here, for example, the coefficient
$ \left( \partial r_+  /\partial Q \right)_{T} $
is computed as follows. From Eq. (\ref{eq:temperature}),
$ r_+ $ and $ q $
are rewritten as
$ r_+  =  r_+ ( T_0, q ) $
and
$ q    =  q   (T_0, r_+)$,
and the charge
$ Q ( r_+, q ) $
can be written as
$ Q = Q( r_+ ( T_0, q ), q)	= Q( r_+,  q( T_0, r_+ ) ) $.
Differentiating these relations with respect to $ r_+ $ and $ q $, we obtain the coefficient.

We find that the correction of $O(\epsilon^2)$ vanishes as for the entropy difference. 
\begin{eqnarray}
	\frac{ \Delta F^{(1)} }{ F_0 }
	 =
	\delta F_1
	 =
	\frac{ 1 }{ 1 + (d-2){\Phi_{H 0}^2} } 
		\Big[
			(d-1){\Phi_{H 0}^2} ~ \delta Q_1-(d-3)(1-{\Phi_{H 0}^2}) ~\delta T_1
		\Big]. 
\end{eqnarray}
Thus, the non-vanishing leading term comes from $O(\epsilon^4)$:
\begin{eqnarray}
 &&
 	\frac{ F_{ \mathrm{NU} } - F_{ \mathrm{U} } }{ F_{ \mathrm{U} } }
	 =
 	- \frac{ (d-3) ( 1 - {\Phi_{H 0}^2} ) }{ 2[ 1 + (d-2){\Phi_{H 0}^2}] } 
	 \biggl\{
  			\frac{ {\Phi_{H 0}^2} (d-1) \delta Q_1 }{ [1-(d-2){\Phi_{H 0}^2} ]} 
    	 	 \left[
    			\frac{ \delta Q_1 }{ d-3 }
    			+ 2 \delta T_1 
    			- \frac{ 1 - (d-2){\Phi_{H 0}^2}}{ (d-3)(1-{\Phi_{H 0}^2}) }\delta \Phi_{H 1}
    		 \right]
    		  \nonumber
    		  \\
  			  && \hspace{7cm}
  		  	 + \delta T_1
   			\left[
   				\delta S_1 + \frac{(d-2)-{\Phi_{H 0}^2}}{1-(d-2){\Phi_{H 0}^2}} \delta T_1
   			\right]
	 \biggr\}
			\epsilon^4.
\end{eqnarray}

%%%%%%%%%%%%%%%%%%%%%%%%%%%%%%%%%%%%%%%%%%%%%%%%%%%%%%%%%%%%%%%%%%%%%%%%%%%%%%%

%%%%%%%%%%%%%%%%%%%%%%%%%%%%%%%%%%%%%%%%%%%%%%%%%%%%%%%%%%%%%%%%%%%%%%%%%%%%%%%
%\bibliographystyle{apsrev} 
\bibliographystyle{unsrt} %% plain, unsrt, alpha, abbrv, acm, apalike
%% \bibliography{BH02.bib}

%%%%%%%%%%%%%%%%%%%%%%%%%%%%%%%%%%%%%%%%%%%%%%%%%%%%%%%%%%%%%%%%%%%%%%%%%%%%%%%

\end{document}